\documentclass{ptephy_v1}

\preprintnumber{XXXX-XXXX} 
\usepackage{amsthm}
\usepackage{bm,latexsym,amsmath,amssymb,amsfonts}
\usepackage{graphics}
\usepackage{ulem}

\begin{document}

\title{Effects of chiral MIT boundary conditions for a Dirac particle in a box}

\author{Ar Rohim$^{1}$ and Kazuhiro Yamamoto$^{2,3}$}
\affil{$^1$Department of Physical Science, Graduate School of Science, Hiroshima University, Higashi-Hiroshima, 739-8526, Japan 
 \\
 $^2$Department of Physics, Kyushu University, 
  Fukuoka,  819-0395, Japan
  \\
 $^3$Research Center for Advanced Particle Physics, Kyushu University, 
 Fukuoka 819-0395, Japan}

\begin{abstract}
We investigate the effects of the chiral MIT boundary conditions for a Dirac particle in a 1D box. We show how the boundary condition affects the discrete momentum, energy level, transition frequency, and spin state in a box. 
The effects of the chiral MIT boundary conditions on the probability and scalar densities are also demonstrated. The results show that an asymmetric distribution appears in the box depending on the parameters of the spin orientation and chiral angle. 
\end{abstract}

\subjectindex{A64, B69}

\maketitle

\section{Introduction}

Particle confinement systems have been intensively studied. 
A basic and well-known example is a free non-relativistic (NR) particle in an infinite potential well. In such systems, it is well known that momentum and energy states become discrete. Studies on particle confinement are not only applicable to NR particles, but also to relativistic particles, such as Klein–Gordon (KG)  \cite{Alberto2017, Alkhateeb}, Dirac \cite{Alberto, Alonso1997, Toyama2010a, Toyama2010, Alberto11, AlHashimi2011, Sitenko2014, Menon2004}, and Majorana particles \cite{AlHashimi} with various shapes or forms of confinement models.

It is also well known that the boundary condition for a spin particle is not trivial. For NR and KG particles, it is sufficient to use Dirichlet boundary conditions, which require the vanishing of the wavefunction and its probability density around the boundary \cite{Alberto2017}. However, for Dirac and Majorana particles, the Dirichlet boundary conditions cannot be applied. This condition inspired the authors of Refs.~\cite{Alberto,Alberto11} to use the boundary condition from the MIT bag model (BC-MIT) \cite{Chodos1,Chodos2, Johnson} when they considered the Dirac particle in a box. BC-MIT is not a Dirichlet boundary condition; it leads to the vanishing of a normal probability current and scalar densities at the boundary surface \cite{AWThomas}. The non-vanishing wavefunction around the boundary surface implies that the axial current is not conserved or the chiral symmetry is broken \cite{AlHashimi2011}. To solve this issue,  the authors of Ref.~\cite{Chodos1975} introduced the chiral bag model (see also, e.g., Refs.~\cite{Brown1979,Hosaka1996}), in which the axial current is preserved or the chiral symmetry is restored by introducing a pion field outside the bag, as has been mentioned in Refs.~\cite{AlHashimi2011,Holt2014}. Here, BC-MIT is not unique as the boundary condition \cite{Theberge1980}. 
In other words, one can also introduce other boundary conditions that guarantee the vanishing of normal probability current density around the boundary surface, e.g., the boundary conditions in Refs.~\cite{AlHashimi2011, Theberge1980, Jaffe, Lutken, AlHashimi2012, AlHashimi2015}. The general form of BC-MIT, which includes the contribution of the chiral angle, is called the chiral MIT boundary conditions (BC-chiral MIT) \cite{Jaffe, Lutken, Theberge1980}. 
Similar to BC-MIT, BC-chiral MIT can also be used as an alternative boundary conditions for the confinement system consisting of a perfectly reflecting mirror. However, under BC-chiral MIT, the energy level of a relativistic particle is allowed to depend on the chiral angle \cite{Jaffe}. BC-chiral MIT has been applied in many studies. For example, the authors of Ref.~\cite{Chernodub} used BC-chiral MIT to confine Dirac fermions in a rotating system; see also Refs.~\cite{Chernodub2016,Chernodub2017, Ambrus2015}. The author of Ref.~\cite{Nicolaevici} showed that the interaction between the particle and the mirror can change the spin orientation depending on the chosen chiral angle.

The above properties of BC-chiral MIT inspired us to revisit the system of a particle confinement in a 1D box consisting of two perfectly reflecting mirrors. In the present paper, we analyze a Dirac particle in Minkowski coordinates under BC-chiral MIT \cite{Lutken}. This study is a generalization of the study in Ref.~\cite{Alberto}, where an analysis of a relativistic particle in a 1D box was conducted using BC-MIT. As shown in Ref.~\cite{Nicolaevici}, the spin orientation of the particle can be changed owing to the interaction between the particle and the boundary surface.  In the system of confinement of a Dirac particle in a box, reflection at the boundary leads to a nontrivial state based on the spin. The stationary state in a box is determined by the change of the spin state at the boundary.

We focus our investigation on the basic properties of wavefunctions of a Dirac field with BC-chiral MIT. This should be useful in studying the finite volume effect associated with Dirac fields. For example, an application for the chiral symmetry breaking of Dirac fermions will be a relevant problem. A study on an such application for the Nambu–Jona-Lasinio  model has been conducted in, e.g., Refs.~\cite{Chernodub2016,Chernodub} (see also Ref.~\cite{Zhang2019} for a work using BC-MIT). In this study, we first construct the wavefunction for a Dirac particle in a 1D box determined by the interaction with the mirrors under BC-chiral MIT, which changes or rotates the spin orientations depending on the chiral angle. 
Using the Dirac wavefunction under BC-chiral MIT, we determine how the energy level depends on the chiral angle. Because BC-chiral MIT is related to the normal probability current and scalar densities, particularly around the boundary surface, it is also essential to include an analysis of these parameters together with the probability density. The properties obtained  for these parameters can be compared with those of the NR, ultra-relativistic (UR), and non-chiral cases \cite{Alberto}.

The remainder of this paper is organized as follows.  In Sect.~\ref{SecWaveFunction}, we consider the solution to a Dirac equation in a 1D box by including the contribution of the spin orientations. In Sect.~\ref{boundaryconditionsandboundstate}, we apply the BC-chiral MIT to the Dirac wavefunction to obtain the bound state, energy level, and transition frequency. In Sect.~\ref{changesspinstates}, we analyze the changes of spin orientation owing to reflection under  BC-chiral MIT.  In Sect.~\ref{probabilitydensity}, we analyze the probability, normal probability current, and scalar densities.
Section \ref{Summary} is devoted to the summary and conclusions. In Appendix \ref{DiracMinkownskiSolution}, we briefly review the Dirac solution in Minkowski coordinates. In this paper, we use $\hbar=c=1$. 

\section{Dirac wavefunction in a 1D box}
\label{SecWaveFunction}

We consider a Dirac particle confined in a 1D box consisting of two perfectly reflecting mirrors. The first mirror is placed at $z=0$, whereas the second mirror is placed at $z=\ell$. We assume that the Dirac particle moves with the momentum $k_3$ along the $z$-axis and the perpendicular momentum is suppressed. Inside the box, the Dirac wavefunction consists of a linear combination of the right-moving wave and the left-moving wave components, which are specifically defined as follows: At the first mirror, the incident and reflected components move in the $-z$- and $+z$-directions, respectively. At the second mirror, the incident and reflected components move in the $+z$- and $-z$-directions, respectively, opposite the component around the first mirror. The incident and reflected components around each mirror are associated with their spin orientations;  see Ref.~\cite{Nicolaevici} for the reflection system.

The total Dirac wavefunction for  a Dirac particle in a 1D box can be  written explicitly as 
\begin{eqnarray}
 \psi_{k_3 s}(z)=  
 B
  \begin{pmatrix}
\xi_R \\
{\sigma_3k_3\over (m+E)}\xi_R
\end{pmatrix}
e^{ik_3z}
+
C 
  \begin{pmatrix}
\xi_L \\
{-\sigma_3k_3\over (m+E)}\xi_L
\end{pmatrix} 
e^{-ik_3z}
,
\label{DiracWavefunctionOneDim}
 \end{eqnarray}
where $B$ and $C$ are complex constants. The letters ``$R$" and ``$L$" correspond to right- and left-moving wave components, respectively. Each component of the Dirac wavefunction $\Psi_{k_3 s}(t,z)\equiv\psi_{k_3 s}(z) e^{-iEt}$ satisfies the Dirac equation (\ref{DiracEquation}), where the two-component spinor $\xi_R$ and $\xi_L$ can be decomposed into
\begin{eqnarray}
 &&\xi_L=
   \begin{pmatrix}
\alpha_L \\
\beta_L
\end{pmatrix},
\hspace{1cm}
 \xi_R=
   \begin{pmatrix}
\alpha_R \\
\beta_R
\end{pmatrix}.
\label{xiIR}
\label{spinrelation}
\end{eqnarray}

In our setup, we first assume that the left-moving wave component has an arbitrary spin orientation. The normalized two-component spinor $\xi_L$ (\ref{xiIR}) requires the condition $|\alpha_L|^2+|\beta_L|^2=1$, 
where the values of the components $\alpha_L$ and $\beta_L$ determine the direction of spin orientation, as shown in Appendix~\ref{spincomponent} for general cases. The values of  components $\alpha_L$ and $\beta_L$ for specific
spin orientations are given as follows: 
$(\alpha_L, \beta_L)=(1,0)$ for the $+z$-direction, 
$(0,1)$ for the $-z$-direction, 
$(1,1)/\sqrt{2}$ for the $+x$-direction,
$(1,-1)/\sqrt{2}$ for the $-x$-direction,
$(1,i)/\sqrt{2}$ for the $+y$-direction, and
$(1,-i)/\sqrt{2}$ for the $-y$-direction. Owing to the interaction with the mirror, the two-component spinor $\xi_L$ can be related to the two-component spinor $\xi_R$. The relation is controlled by a rotation operator in the spin space \cite{Nicolaevici}.

\section{Bound states, energy level, and transition frequency}
\label{boundaryconditionsandboundstate}
In this section, we start with a review of BC-chiral MIT \cite{Lutken, Jaffe, Theberge1980}. Then,  we follow the procedure in Ref.~\cite{Alberto} to find the bound states, which makes the allowed momentum discrete. The obtained discrete momentum enables us to investigate the energy level and transition frequency. 
 In the system of a Dirac particle in 1+1D, one can develop a formulation on the basis of $2\times 2$ matrices. This analysis can be found, e.g., in Refs.~\cite{Jaffe, Alonso1997}. By using this formalism for the $1+1$D system, one is able to obtain a clear discussion, e.g., on the NR limit of boundary conditions \cite{Alonso1997}, which is useful to analyze the bound states. Furthermore,  it is applicable not only to the bound state's analysis but also to the spin orientation. However, in the present paper, we develop a formulation on the basis of $4\times 4$ gamma matrices. This is because the formulation can be easily extended to analyses of the problems in a 3D box (see, e.g., Ref.~\cite{Alberto11}) and other general cases.

 \subsection{Boundary conditions: BC-chiral MIT}
 \label{boundaryconditions}

BC-chiral MIT at a mirror is given by \cite{Lutken}
 \begin{eqnarray}
	iN_\mu\gamma^\mu\psi
	=e^{-i\Theta\gamma^5}\psi,
	\label{chiralMITBC}
\end{eqnarray}
where $N_\mu$ is an inward normal unit vector
to the boundary, $\Theta$ is the chiral angle that takes values with the range  $0\leq\Theta < 2\pi$, $e^{-i\Theta\gamma^5}\equiv\cos\Theta I -i\sin\Theta\gamma^5$, and $\gamma^5=i\gamma^0\gamma^1\gamma^2\gamma^3$. Here, we work in the Dirac representation (\ref{DiracRepresentation}). For the non-chiral case $(\Theta=0)$, the above  BC-chiral MIT (\ref{chiralMITBC})  reduces to BC-MIT in a straightforward manner. BC-chiral MIT guarantees a vanishing normal probability current at the boundaries for any chiral angle, which can be understood as follows \cite{Jaffe}: Multiplying both sides of BC-chiral MIT (\ref{chiralMITBC}) with $\bar\psi(\equiv\psi^\dagger \gamma^0)$ 
from the left, we have
\begin{eqnarray}
i N_\mu\bar\psi\gamma^\mu\psi=\bar\psi e^{-i\Theta\gamma^5}\psi.
\label{BC1}
\end{eqnarray}
Taking the Hermitian conjugate of Eq.~(\ref{chiralMITBC}) and then multiplying both sides 
with $\gamma^0\psi$ from the right, we have
\begin{eqnarray}
-i N_\mu\bar\psi\gamma^\mu\psi=\bar\psi e^{-i\Theta\gamma^5}\psi,
\label{BC2}
\end{eqnarray}
where we have  used the anti-commutation relation $\lbrace\gamma^5,\gamma^0\rbrace=0$.
At the boundaries, from Eqs.~(\ref{BC1}) and (\ref{BC2}), we have
\begin{eqnarray}
iJ_N=iN_\mu\bar\psi\gamma^\mu\psi
=0 ~~{\rm and }~~\bar\psi e^{-i\Theta\gamma^5}\psi=0,
\label{probandscalardensities}
\end{eqnarray}
 where $J_N$ denotes the normal probability current density.  The condition  $\bar\psi e^{-i\Theta\gamma^5}\psi=0$ indicates that  BC-chiral MIT guarantees the vanishing scalar density $\bar\psi\psi$  at the boundaries for the  chiral angle $\Theta=0, \pi$ only. 
 Later, in Sec.~\ref{probabilitydensity}, we consider these parameters together with the probability density in more detail.

\subsection{Bound states} 

We consider the bound states of a Dirac particle in a 1D box. 
At the first mirror, the inward normal unit vector to the boundary is given by 
$N_\mu=(0,0,0,1)$.
Applying the BC-chiral MIT (\ref{chiralMITBC}) 
to the Dirac wavefunction  $\psi=(\Phi_1,\Phi_2)^{\rm T}$ gives the following equations:
\begin{eqnarray}
&&i(\sigma_3+\sin\Theta I)\Phi_2|_{z=0}-\cos\Theta\Phi_1|_{z=0}=0,
\label{chiral1}\\
&&i(\sigma_3-\sin\Theta I)\Phi_1|_{z=0}+\cos\Theta\Phi_2|_{z=0}=0.
\label{chiral2}
\end{eqnarray}
Both the boundary conditions in Eqs. (\ref{chiral1}) and (\ref{chiral2}) are equivalent. This can be understood as follows: from Eq.~(\ref{chiral1}) we have $\Phi_1=i(\sigma_3+\sin\Theta I)\Phi_2/ \cos\Theta$. This $\Phi_1$ trivially satisfies the boundary condition (\ref{chiral2}). This condition implies that one can freely choose one of them \cite{Nicolaevici}. 

Imposing the boundary condition (\ref{chiral1}) on the Dirac wavefunction (\ref{DiracWavefunctionOneDim}) at the first mirror $z=0$, we have  
 \begin{eqnarray}
 B \left[i(I+\sin\Theta\sigma_3) {k_{3}\over (m+E)}-\cos\Theta I\right]\xi_R= C \left[i(I+\sin\Theta\sigma_3) {k_{3}\over (m+E)}+\cos\Theta I\right]\xi_L.
 \label{BCrotationangle0}
  \end{eqnarray}
  The relation above can be rewritten explicitly as
\begin{eqnarray}
 \begin{pmatrix}
\left[i(1+\sin\Theta){k_3\over (m+E)}-\cos\Theta\right]\alpha_R~~ &~~ \left[-i(1+\sin\Theta){k_3\over (m+E)}-\cos\Theta \right]\alpha_L\\
\left[i(1-\sin\Theta){k_3\over (m+E)}-\cos\Theta\right]\beta_R ~~&~ \left[-i(1-\sin\Theta){k_3\over (m+E)}-\cos\Theta\right]\beta_L
\end{pmatrix}
 \begin{pmatrix}
B\\
C
\end{pmatrix}
=0,
\label{MatrixrelationBC}
\end{eqnarray}
where we use the expression of the two-component spinor, i.e., Eq.~(\ref{spinrelation}). The values of coefficients $B$ and $C$ cannot be zero because if one takes one or both of them to be zero, the Dirac wavefunction (\ref{DiracWavefunctionOneDim}) vanishes everywhere. Thus, we require a 
vanishing of the determinant of the $2\times 2$ matrix in Eq.~(\ref{MatrixrelationBC}).
This requirement implies that the components of the incident and reflected spin 
orientations satisfy the following equation:
\begin{eqnarray}
(\alpha_R\beta_L-\beta_R\alpha_L)-{ik_3\tan\Theta\over E}(\alpha_R\beta_L+\beta_R\alpha_L)=0. 
\label{detmatrix}
\end{eqnarray}
Equation~(\ref{detmatrix}) shows that the reflected spin orientation depends on the chiral angle, mass, momentum, and incident spin orientation, as  will be shown in Sect.~\ref{changesspinstates}. From Eq.~(\ref{MatrixrelationBC}), we have two equations denoting the relations between coefficients $B$ and $C$ as
\begin{eqnarray}
&&C=B
{i (1+\sin\Theta) k_3/(m+E)-\cos\Theta \over i (1+\sin\Theta) k_3/(m+E) + \cos\Theta} {\alpha_R\over \alpha_L},
\label{CoefficientRelations1}\\
&&C=B
{i (1-\sin\Theta) k_3/(m+E)-\cos\Theta \over i (1-\sin\Theta) k_3/(m+E) + \cos\Theta} {\beta_R\over \beta_L}.
\label{CoefficientRelations2}
\end{eqnarray}
Equation (\ref{CoefficientRelations1}) connects the relation between $\alpha_R$ with $\alpha_L$, whereas Eq.~(\ref{CoefficientRelations2}) connects  $\beta_R$ with $\beta_L$. If the spin orientation does not change with the chiral angle $\Theta=0,\pi$, they yield the same equation. The reflected spin orientation satisfies Eq.~(\ref{detmatrix}), which yields Eqs.~(\ref{CoefficientRelations1}) and (\ref{CoefficientRelations2}) as essentially equivalent. We use these relationships in the next analysis.

To find the expression of the discrete momentum, we next proceed to BC-chiral MIT at the second mirror, where the inward normal unit vector to the boundary is given by $N_\mu=(0,0,0,-1)$.  Then, BC-chiral MIT for the Dirac wavefunction $\psi=(\Phi_1,\Phi_2)^{\rm T}$ reads,
\begin{eqnarray}
&&i(\sigma_3-\sin\Theta  I)\Phi_2|_{z=\ell}+\cos\Theta\Phi_1|_{z=\ell}=0,
\label{chiral1L}\\
&&i(\sigma_3+\sin\Theta  I)\Phi_1|_{z=\ell}-\cos\Theta\Phi_2|_{z=\ell}=0.
\label{chiral2L}
\end{eqnarray}
Similar to the boundary conditions at the first mirror,  Eqs.~(\ref{chiral1L}) and (\ref{chiral2L})  are also equivalent.

\begin{figure}[t]
\centering\includegraphics[width=4.2in]{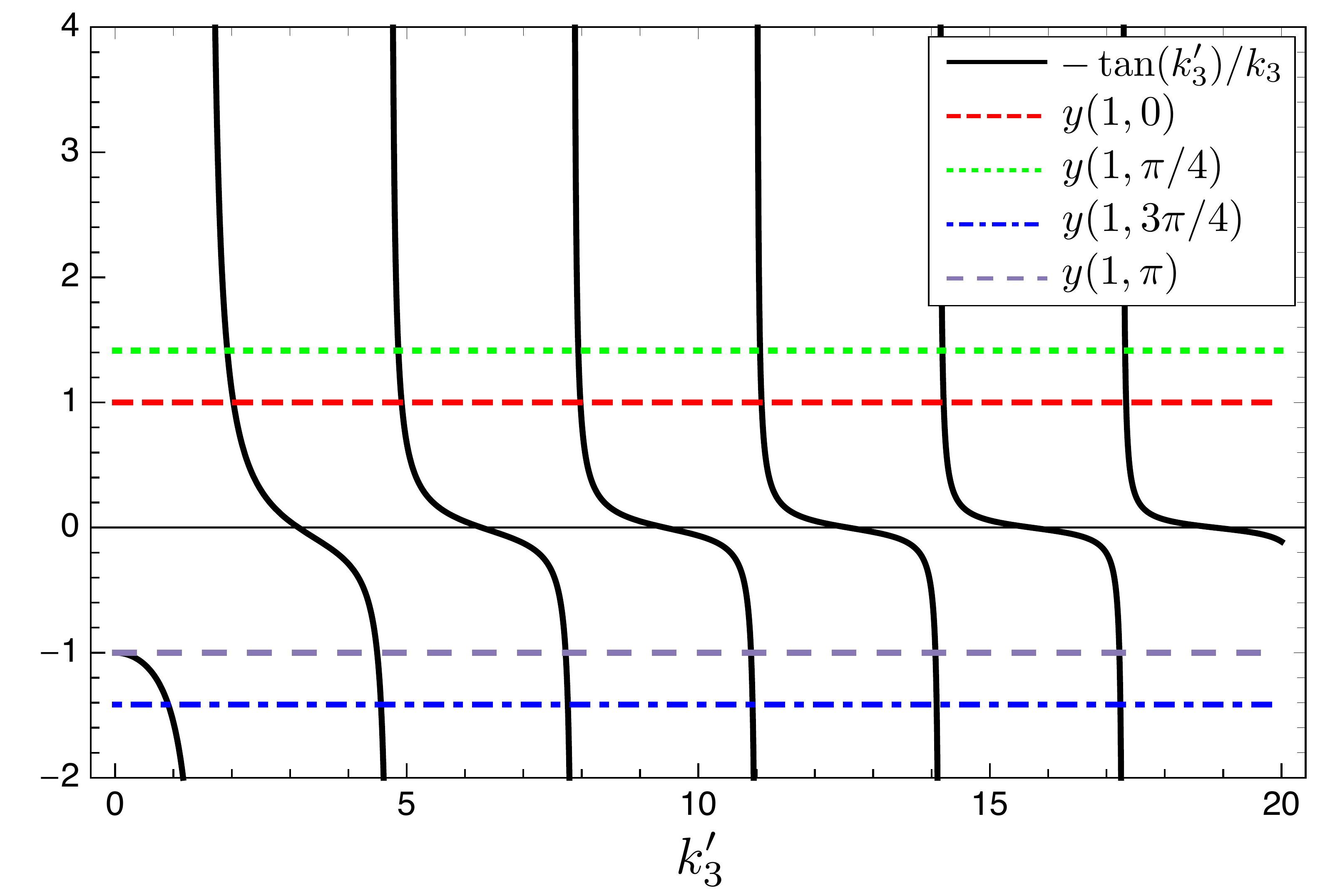}
\caption{ Plot of $-\tan(k^\prime_3)/k^\prime_3$ and $1/m'\cos\Theta\equiv y(m',\Theta)$ as a function of $k_3'$.
The intersection between the curve $-\tan(k^\prime_3)/k'_3$ and the horizontal line  $y(m',\Theta)$ gives the solutions of the discrete momentum. Herein,  we adopt $m'=1$ and $\Theta=0,\pi/4,\pi,3\pi/4$. 
\label{EnergyLevelOneDim}}
\end{figure}

\begin{figure}[t]
\centering\includegraphics[width=4.2in]{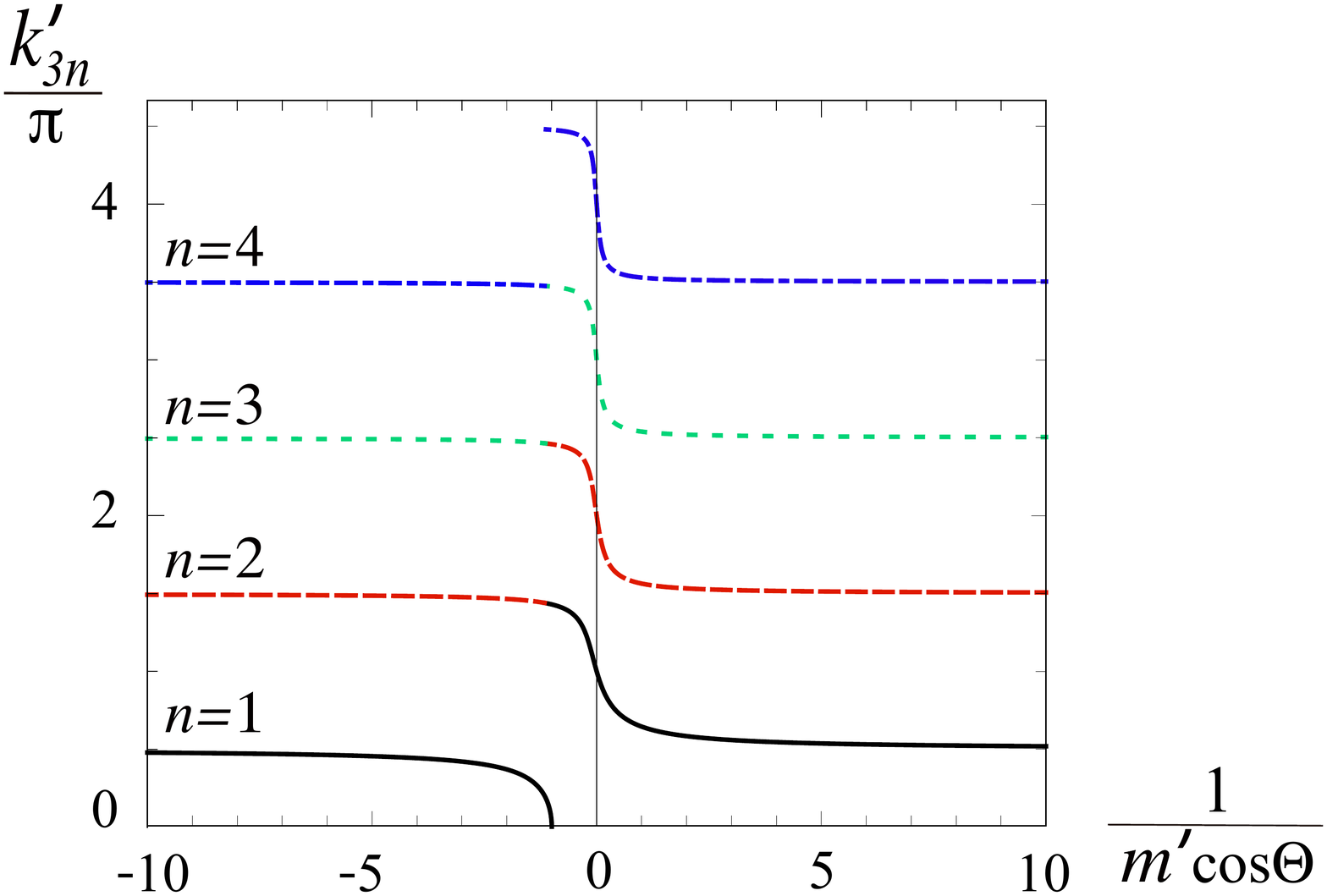}
\caption{Solution $k_{3n}'$ for Eq.~(\ref{discretek32}), 
normalized by $\pi$ as a function of  $1/m'\cos\Theta$. The black solid curve, the red dashed curve, the green dotted curve, the blue dash-dotted curve are the solutions to $n=1,~2,~3,~4$, respectively.}
\label{EnergyLevelOneDim2}
\end{figure}

Imposing the boundary condition (\ref{chiral1L}) on the Dirac wavefunction (\ref{DiracWavefunctionOneDim}), we obtain the following equation:
\begin{eqnarray}
B e^{ik_3 \ell} \left[i(I-\sin\Theta\sigma_3){k_3\over (m+E)}+\cos\Theta I\right]\xi_R
~~~~~~~~~~~~\nonumber\\
~~~~~~~~~~~~
 =C e^{-ik_3 \ell}\left[i(I-\sin\Theta\sigma_3){k_3\over (m+E)}-\cos\Theta I\right]\xi_L.
 \label{xiRL2}
\end{eqnarray}
Using the components in Eq.~(\ref{xiIR}), the above relation (\ref{xiRL2}) can be decomposed into two equations as follows:
\begin{eqnarray}
B e^{ik_3 \ell} \left[i(1-\sin\Theta){k_3\over (m+E)}+\cos\Theta\right]\alpha_R
~~~~~~~~~~~~\nonumber\\
~~~~~~~~~~~~
=C e^{-ik_3 \ell}\left[i(1-\sin\Theta){k_3\over (m+E)}-\cos\Theta\right]\alpha_L,
 \label{BCL1}
 \\
B e^{ik_3 \ell} \left[i(1+\sin\Theta){k_3\over (m+E)}+\cos\Theta\right]\beta_R
~~~~~~~~~~~~\nonumber\\
~~~~~~~~~~~~
=C e^{-ik_3 \ell}\left[i(1+\sin\Theta){k_3\over (m+E)}-\cos\Theta\right]\beta_L.
  \label{BCL2}
 \end{eqnarray}
Substituting the relation (\ref{CoefficientRelations1}) into Eq.~(\ref{BCL1}), we arrive at the following equation for the momentum $k_3$:
\begin{eqnarray}
\tan (k_3 \ell)=-{ k_3\over m \cos\Theta}~,
\label{discretek31}
\end{eqnarray}
which makes the allowed momenta discrete. The same expression is obtained when we substitute the coefficient relation (\ref{CoefficientRelations2}) into Eq.~(\ref{BCL2}). The solutions of discrete momentum (\ref{discretek31}) do not depend on spin orientation because the contribution of spin orientation can be factorized out. After substituting Eqs.~(\ref{CoefficientRelations1}) and (\ref{CoefficientRelations2}) into Eqs.~(\ref{BCL1}) and (\ref{BCL2}), respectively, both the left- and right-hand sides of Eq.~(\ref{BCL1})  
 contain either $\alpha_L$ or $\alpha_R$,  whereas those of Eq.~(\ref{BCL2})  
contain either $\beta_L$  or $\beta_R$. Introducing the new parameters 
\begin{eqnarray}
 && m^\prime=m \ell,
\label{mprime}
\\
&&k_3^\prime=k_3 \ell~, 
\end{eqnarray}
Eq.~(\ref{discretek31}) can be rewritten as
\begin{eqnarray}
-{\tan (k_3^\prime)\over k_3^\prime}={ 1\over m^\prime\cos\Theta}~. 
\label{discretek32}
\end{eqnarray}
It can be seen that the solution for $k^\prime_3$ depends on $m^\prime(=m \ell)$ and chiral angle $\Theta$.
This means that the solution depends on mass $m$, the distance between two mirrors $\ell$, and the chiral angle $\Theta$, but it does not depend on the spin orientation, as noted above.
Figure~\ref{EnergyLevelOneDim} plots the curves of $-\tan k_3'/k_3'$ together with the line $1/m'\cos\Theta$ as a function of $k_3^\prime$. Figure~\ref{EnergyLevelOneDim2} plots the solution of the positive momentum $k_{3n}'$ with $n=1,~2,~3,~4$ as functions of $1/m'\cos\Theta$. 

Let us consider solutions with the discrete momenta for some limits and specific values of the chiral angle. As we demonstrate below, the solutions do not depend on the chiral angle in the significantly UR and NR limits. 
In the UR limit $m'\ll1$, we have an approximate solution for Eq.~(\ref{discretek32})
\begin{eqnarray}
k^{\prime {\rm UR}}_{3n}
\simeq{(2n-1)\pi\over 2}+{2m'\cos\Theta\over (2n-1)\pi}, ~~n=1,2,3, ...
\label{UR0}
\end{eqnarray}
 Note that we here use the notation $n=1,2,3,...$ to denote  ground, first excited, second excited and additional states. Only the first term on the right-hand side of Eq.~(\ref{UR0}) remains irrespective of the mass when we take the chiral angle $\Theta=\pi/2,3\pi/2$. For the NR limit 
 $m^\prime\gg1$,  we obtain the approximate solution
\begin{eqnarray}
k^{\prime{\rm NR}}_{3n}\simeq n\pi\left(1-{1\over m'\cos\Theta}\right).
\label{K3NR}
\end{eqnarray}
For the non-chiral case, the solution (\ref{discretek31}) is reduced to the solution in  Ref.~\cite{Alberto}. This condition corresponds to BC-MIT.

Note that the allowed momentum $k_{3n}'$, for the case of $m'\cos\Theta<0$,  is  lower (for $1/m'\cos\Theta\leq-1$)
and higher (for $-1< 1/m'\cos\Theta<0$) than the values of the UR limit. In Table~\ref{tablek3oneDim}, we provide numerical values of the solutions $k'_{3n}$ for the parameters  $m'=0.5,5$ and chiral angle $\Theta=0,\pi/2,3\pi/4$. We can see that the solutions $k'_{3n}$ depend on the values of  $m'$ and the chiral angle $\Theta$ in general. However, for the chiral angle $\Theta=\pi/2,3\pi/2$, the solution of discrete momentum does not depend on the parameter  $m'$. 

\begin{table}[b]
\centering
\begin{tabular}{|c| |c|  |c| |c| |c| |c|}
  \hline
  $n$ & $k_{3n}^\prime(0.5, 0)$ 
   & $k^\prime_{3n} (m', \pi/2)$ & $k^\prime_{3n}(0.5,3\pi/4)$& $k_{3n}^\prime(5, 0)$ 
   &  $k^\prime_{3n}(5,3\pi/4)$\\
  \hline
   1 & 1.837   &  1.571    & 1.306  &2.654  & 3.987 \\ 
   2 & 4.816   &  4.712    & 4.636  & 5.454 &  7.409\\
   3 & 7.917   &  7.854    &  7.809 & 8.391  & 10.676 \\ 
   4 & 11.041  &  10.996  & 10.963  & 11.409 & 13.888\\ 
   5 & 14.172  & 14.137   & 14.112  & 14.470 & 17.075 \\
   \hline
\end{tabular}
\caption{The lowest five solutions of the discrete momentum $k^\prime_{3n} (m', \Theta)$ for $m'=0.5, 5$  and chiral angle $\Theta=0,\pi/2, 3\pi/4$.\label{tablek3oneDim}} 
\vspace{5mm}
\begin{tabular}{|c||c||c||c||c||c||c|}
  \hline
  $n$ & $E^\prime_{ n}(0.5,0)$ 
  &  $E^\prime_{ n} (0.5,\pi/2)$ & $E^\prime_{ n}(0.5,3\pi/4) $&  $E^\prime_{ n} (5,0)$ & $E^\prime_{ n} (5,\pi/2)$ &$E^\prime_{ n}(5,3\pi/4)$ 
 \\
  \hline
   1 & 1.904 &  1.649 & 1.398  & 5.661 & 5.241 & 6.395\\ 
   2 & 4.842 &  4.738 & 4.663  & 7.399 & 6.870 & 8.938 \\
   3 & 7.933 &  7.870 & 7.825 & 9.768 & 9.310 & 11.789\\ 
   4 & 11.052 &  11.007 & 10.974 & 12.456 & 12.079 & 14.761\\ 
   5 & 14.181 & 14.146 & 14.121 & 15.310& 14.995 & 17.792 \\
   \hline
\end{tabular}
\caption{The lowest five of the energy levels $E'_{n}(m',\Theta)$ for $m'=0.5, 5$  and chiral angle $\Theta=0,\pi/2, 3\pi/4$.
\label{EnergyoneDim}}
\vspace{5mm}
\centering
\begin{tabular}{|c||c||c||c||c||c||c|}
  \hline
  $n$ & $\omega'_p(0.5, 0)$ 
    & $\omega'_p(0.5, \pi/2)$ & $\omega'_p (0.5, 3\pi/4)$ & $\omega'_p(5, 0)$ 
    & $\omega'_p(5, \pi/2)$ & $\omega'_p(5, 3\pi/4)$ 
  \\
  \hline
   1 & 2.938  &  3.089  & 3.265  & 1.738 & 1.629 & 2.543\\ 
   2 & 3.091  &  3.132  & 3.162  & 2.369 & 2.440 & 2.851\\
   3 & 3.119  &  3.137  & 3.149 & 2.688 & 2.769 & 2.972 \\ 
   4 & 3.129  &  3.139  & 3.147  &2.854 & 2.916 & 3.031\\ 
      \hline
\end{tabular}
\caption{Transition frequencies of the few lowest of $\omega^\prime_p(m', \Theta)$ for $m'=0.5, 5$  and chiral angle $\Theta=0,\pi/2,3\pi/4$. 
Here, $p\equiv n+1,n$.
\label{transitionfrequenciesoneDim}}
\end{table}

From Eqs.~(\ref{UR0}) and (\ref{K3NR}), it can be seen that the momentum is discretized in different ways between the UR limit and the NR limit depending on the box's size $\ell$ and the parameter $m\cos\Theta$, where the mass $m$ specifies the Compton wavelength \cite{Alberto, Alberto11}. The correction term of the UR limit decreases as the state $n$ increases, while the correction term of the NR limit increases as the states $n$ increases. However, the contribution of the correction term is relatively small compared with the first dominant term in both the limits. Then, one can say that the discrete momentum does not depend on the chiral angle in either the UR or NR limits. Thus the discrete momentum with BC-chiral MIT is shared with the result with BC-MIT in Refs.~\cite{Alberto, Alberto11} in the UR and NR limits.  Taking the NR limit of the Dirac wavefunction (\ref{DiracWavefunctionOneDim}), BC-chiral MIT reduces to the Dirichlet boundary condition, which requires that the wavefunction vanishes at both boundaries $z'=0$ and $z'=1$ after omitting the contribution of ${\cal O}(1/m')$ and its higher order. In this condition, the discrete momentum is given by the first term of Eq.~(\ref{K3NR}), which is the same as the trivial solution of discrete momentum for the Schr\"{o}dinger equation. However, when we take the UR limit of  the Dirac wavefunction (\ref{DiracWavefunctionOneDim}), BC-chiral MIT does not reduce to the Dirichlet boundary conditions. In this limit, the chiral angle still plays a role in the boundary conditions, where such a property does not appear in the NR limit. The boundary conditions dependence on the chiral angle generates the spin orientation dependence on the chiral angle, as will be shown later in Sect.~\ref{changesspinstates}. To explain the physical meaning further, let us revisit the non-chiral case \cite{Alberto, Alberto11}, where the different properties between the UR and NR limits can also be seen from the relation between the coefficients of the right-moving wave $B$ and left-moving wave $C$. For the reflection at the first mirror, the relation in the NR limit is given by $B=-C$, while in the UR limit, the relation is given by $B=-iC$ though both relations satisfy $|B|^2=|C|^2$. The relation generated by the reflection at the second mirror between the UR and NR limits is also different, namely, $B e^{ik_3\ell}=iCe^{-ik_3\ell}$ and $Be^{ik_3\ell}=-Ce^{-ik_3\ell}$ for the UR and NR limits, respectively. By combining the reflections at both mirrors, one finds that the NR and UR limits give the condition for discrete momentum as $\sin(k_3 \ell)=0$ and $\cos(k_3 \ell)=0$, respectively. Then, finally, we recover the results in the first term of Eqs.~(\ref{UR0}) and (\ref{K3NR}) for the UR and NR limits, respectively. 

From this description, one can see that the boundary conditions for the different limits of the Dirac wavefunction generate different solutions for the discrete momenta. The same scenario also happens for the analysis using BC-chiral MIT. In the NR limit, omitting the contribution of ${\cal O}(1/m')$ and its higher order, the spin orientation does not change and then the reflection at the first and second mirrors under BC-chiral MIT is completely the same as that with BC-MIT. For the UR limit, the situation is not trivial because the presence of a chiral angle in the UR limit changes the spin orientation, as we will show later, in Sect.~\ref{changesspinstates}. In this limit, the reflections at the first and second mirrors give $B\big[i(I+\sin\Theta\sigma_3)-\cos\Theta I\big]\xi_R=C\big[i(I+\sin\Theta\sigma_3)+\cos\Theta I\big]\xi_L$ and $Be^{ik_3\ell}\big[i(I-\sin\Theta\sigma_3)+\cos\Theta I\big]\xi_R=Ce^{-ik_3\ell}\big[i(I-\sin\Theta \sigma_3)-\cos\Theta I\big]\xi_L$, respectively. Interestingly, in this limit, we arrive at the condition for discrete momentum  as $\cos(k_3\ell)=0$, which is the same as BC-MIT. These results show that the chiral angle does not affect the solution of discrete momentum in the NR and UR limits, but it will affect the spin orientation. To see the contributions of the chiral angle, we must at least include the order of ${\cal O}(1/m')$ and ${\cal O}(m')$ for NR and UR limits, respectively. Compared to BC-MIT, the differences generated by BC-chiral MIT appear in the second terms of Eqs.~(\ref{UR0}) and (\ref{K3NR}) for the UR and NR limits, respectively. From BC-chiral MIT framework, one can understand that a different limit generates a different discrete momenta solution, similar to that of BC-MIT.

\subsection{Energy level}

In the above derivation, we have shown that BC-chiral MIT implies that the momentum of the Dirac particle in a box becomes discrete, the values of which
depend on the parameters $m^\prime(=m\ell)$ and chiral angle $\Theta$. The discrete momentum leads to the energy level, which can be written as
\begin{eqnarray}
E^\prime_n =\sqrt{m^{\prime2}+k^{\prime2}_{3n}},
\label{EnergyLevel1}
\end{eqnarray}
where $E^\prime_n=E_n \ell$ and the solution of momentum $k^\prime_{3n}$ 
satisfies Eq.~(\ref{discretek32}). 

Similar to the discrete momentum, the energy level depends on $m'$ and chiral angle $\Theta$, but does not depend on the spin orientation. In Table~\ref{EnergyoneDim}, we provide the energy levels of a Dirac particle for $m'=0.5, 5$ and three values of chiral angle $\Theta=0,\pi/2,3\pi/4$, which 
correspond to the values in Table~\ref{tablek3oneDim}.

Let us consider the UR and NR limits for the energy level (\ref{EnergyLevel1}). In the UR limit $m^\prime\ll 1$, the energy level is reduced to
\begin{eqnarray}
E^{\prime {\rm UR}}_n \approx k^{\prime {\rm UR}}_{3n}\simeq{{(2n-1)\pi\over 2}+{2m'\cos\Theta\over (2n-1)\pi}},
\end{eqnarray}
after omitting the higher-order terms of ${\cal O}(m^{\prime 2})$. In this limit, because the values of $m'$ are quite small, we can see that the energy level does not depend on the chiral angle $\Theta$.  
In the NR limit $m^\prime \gg 1$, we have the approximate form of the energy level as
\begin{eqnarray}
E^{\prime {\rm NR}}_n \approx m'+{(k^{\prime {\rm NR}}_{3n})^2\over 2m'}\simeq m'+{(n\pi)^2\over 2m'},
\label{EnergyLevelNR}
\end{eqnarray}
after omitting the higher order of  ${\cal O}({1/m'^2})$.  Note that the first term in Eq.~(\ref{EnergyLevelNR}) is equivalent to the rest mass $m \ell$, which means that the energy level only depends on the rest mass of the Dirac particle and that the dependence on the chiral angle is small because it becomes the higher-order corrections. In other words, we can say that the energy level in the NR limit also does not depend on the chiral angle.

\subsection{Transition frequency}

The transition frequency between energy eigenstates $E'_n$ and $E'_{n+1}$ is explicitly given by (cf. Ref.~\cite{Jenke2011}):
\begin{eqnarray}
\omega^\prime_{n+1,n}&=&E^\prime_{n+1}-E^\prime_n,
\end{eqnarray}
 where the energy level is given by Eq.~(\ref{EnergyLevel1}), with  a discrete momentum satisfying Eq.~(\ref{discretek32}). Similar to its discrete momentum and energy level, the transition frequency also  depends on $m'$ and chiral angle $\Theta$, but does not depend on the spin orientation. Compared to the non-chiral case, the values of the transition frequency can be lower or higher depending on the chiral angle. In Table.~\ref{transitionfrequenciesoneDim}, we provide the transition frequencies for the few lowest states. 

In the UR limit, we obtain an approximate solution for the transition frequency, which is explicitly given as
\begin{eqnarray}
 \omega^{\prime{\rm UR}}_{n+1,n}\simeq\pi-{4m'\cos\Theta\over (4n^2-1)\pi}. 
 \label{TFUR}
\end{eqnarray}
In this limit, the transition frequency for all chiral angles converges to $\pi$ as the state $n$ increases. In the NR limit, the transition frequency is reduced to the following:
\begin{eqnarray}
 \omega^{\prime{\rm NR}}_{n+1,n}\simeq{(2n+1)\pi^2\over 2m'}.
 \label{TFNR}
\end{eqnarray}
In contrast with the UR limit, the transition frequency for the NR limit increases as the energy level $n$ increases. The approximate expressions in Eqs.~(\ref{TFUR}) and (\ref{TFNR}) show that the dependence on the chiral angle $\Theta$ is the higher-order corrections, which is similar to the behavior of the discrete momentum and energy level for the same limits. 
 
\section{Changes of spin orientation}
\label{changesspinstates}

In this section, we analyze the changes of spin orientation generated by the reflection or the interaction with the mirrors. As shown in Ref.~\cite{Nicolaevici}, the change is determined by the rotation operator in the spin space.  First, we review the changes of spin orientation generated by the reflection at the first mirror following the procedure in Ref.~\cite{Nicolaevici}, but with suppressed perpendicular momentum and with the boundary condition (\ref{chiralMITBC}) as provided in Ref.~\cite{Lutken}. Using the same procedure, we analyze the 
changes of spin orientation generated by the reflection at the second mirror. At the end of this section, we describe the relationship between the changes of spin orientation in both mirrors.

\subsection{Reflection at first mirror}
At the first mirror, we have the relation between the two-component spinor $\xi_L$ and $\xi_R$ as given in Eq.~(\ref{BCrotationangle0}).  To proceed, we follow the procedure in Ref.~\cite{Nicolaevici}. It is more convenient to use the following parameters:
\begin{eqnarray} 
&&Q^{(1)}_L=C\left[i(I+\sin\Theta\sigma_3) { k'_{3n}\over (m'+E'_n)}+\cos\Theta I\right]
  \label{Q0I}
,
\\
&&Q^{(1)}_R=B\left[i(I+\sin\Theta\sigma_3) {k'_{3n}\over (m'+E'_n)}-\cos\Theta I\right]
\label{Q0R}
.
\end{eqnarray}
Then, the relationship between incident and reflected spin orientations around the first mirror (\ref{BCrotationangle0}) reads
  \begin{eqnarray}
  \xi_R=(Q^{(1)}_R)^{-1}Q^{(1)}_L\xi_L={\cal U}^{(1)}\xi_L,
  \label{xi0Rxi0I}
  \end{eqnarray}
  where ${\cal U}^{(1)}$ represents the rotation operator in a spin space, which can be written as follows \cite{Nicolaevici}:
  \begin{eqnarray}
  {\cal U}^{(1)}=e^{i\chi^{(1)}}\left[\cos({\varphi^{(1)}_{n}\over 2})I-i\sin({\varphi^{(1)}_{n}\over 2})
  {\bm \Upsilon}^{(1)}\cdot{\bm \sigma}\right].
  \label{sincosrotation0}
  \end{eqnarray}
Here,  $e^{i\chi^{(1)}}$ is a pure phase, $\varphi^{(1)}_{n}$ denotes the rotation angle, and $
{\bm \Upsilon}^{(1)}$ denotes  the unit rotation axis generated by the reflection at the first mirror. Substituting Eqs.~$(\ref{Q0I})$ and $(\ref{Q0R})$ into Eq.~(\ref{xi0Rxi0I}), the relation between the incident and reflected spin orientations at the first mirror can be written as
\begin{eqnarray}
\xi_R= {C\over B} {\cos\Theta\over(ik'_{3n}-m'\cos\Theta)}
\left( E'_n I+ik'_{3n}\tan\Theta\sigma_3\right)
\xi_L
\label{spinincidentreflected0}
.
\end{eqnarray}
Using the two-component spinor as provided in Eq.~(\ref{xiIR}), 
Eq.~(\ref{spinincidentreflected0}) can be written in an explicit form as
\begin{eqnarray}
 \begin{pmatrix}
\alpha_R\\
\beta_R
\end{pmatrix}
={C\over B} {\cos\Theta\over(ik'_{3n}-m'\cos\Theta)}
 \begin{pmatrix}
(E'_n+ik'_{3n}\tan\Theta)\alpha_L\\
(E'_n-ik'_{3n}\tan\Theta)\beta_L 
\label{rotationspin}
\end{pmatrix}.
\end{eqnarray}
Using the equivalent expressions in Eq.~(\ref{xi0Rxi0I}) with Eqs.~(\ref{sincosrotation0}) and (\ref{spinincidentreflected0}), we obtain
\begin{eqnarray}
&&e^{i\chi^{(1)}}\cos({\varphi^{(1)}_{n}\over 2})={C \over B} {E'_n\cos\Theta\over(ik'_{3n}-m'\cos\Theta )},\\
&&e^{i\chi^{(1)}}\sin({\varphi^{(1)}_{n}\over 2})
{\bm \Upsilon}^{(1)}\cdot{\bm \sigma}=-{C \over B} {k'_{3n}\sin\Theta\over(ik'_{3n}-m'\cos\Theta)}\sigma_3.
\end{eqnarray}
Finally, we obtain the rotation angle and rotation axis at the first mirror as
\begin{eqnarray}
  \tan({\varphi^{(1)}_{n}\over 2})=-{k'_{3n}\tan\Theta\over E'_n},~~{\rm and }~~~
  {\bm \Upsilon}^{(1)}= \hat z
  \label{rotationangle0},
  \end{eqnarray}
where $\hat z$ is a unit vector in the direction of the $z$-axis.\footnote{We may choose ${\bm \Upsilon}^{(1)}=-\hat z$ but with the opposite sign of $\varphi^{(1)}_{n}$.} 

Because we consider the momentum perpendicular to the 
mirrors, our situation is limited  compared with the situation in Ref.~\cite{Nicolaevici};
however, there are relations with those in Ref.~\cite{Nicolaevici}.
The sign of the rotation angle in Eq.~(\ref{rotationangle0}) is opposite to that of Ref.~\cite{Nicolaevici}, though the rotation axis is the same. The opposite sign of the rotation angle comes from the opposite sign of the definition of the chiral angle $\Theta$, for which we followed the Ref.~\cite{Lutken}. In the present study, we impose another boundary condition at $z'=1$ to find the states in the confinement system of the 1D box.  As we will see in the next subsection, we obtain the same rotation angle with the opposite rotation axis at the boundary $z'=1$. 

\subsection{Reflection at second mirror}
\label{Reflec2ndMirror}

From Eq.~(\ref{xiRL2}), we obtain the relation between the two-component spinor $\xi_L$ and $\xi_R$ at the second mirror. It is more convenient to use the following parameters: 
  \begin{eqnarray}
    &&Q^{(2)}_L=C e^{-ik'_{3n}}\left[i(I-\sin\Theta\sigma_3){k'_{3n}\over (m'+E'_n)}-\cos\Theta I\right]
  ,
  \label{QIL}
\\
    &&Q^{(2)}_R=B e^{ik'_{3n}} \left[i(I-\sin\Theta\sigma_3){k'_{3n}\over (m'+E'_n)}+\cos\Theta I\right]
  .
    \label{QRL}
  \end{eqnarray}
  Then, the relation between the incident and reflected spin orientations around the second mirror  (\ref{xiRL2}) reads
  \begin{eqnarray}
  \xi_L=(Q^{(2)}_L)^{-1}Q^{(2)}_R\xi_R={\cal U}^{(2)}\xi_R,
  \label{xiRL}
  \end{eqnarray}
  where ${\cal U}^{(2)}$ is the rotation operator, which can be written as follows \cite{Nicolaevici}
  :
  \begin{eqnarray}
  {\cal U}^{(2)}=e^{i\chi^{(2)}}\left[\cos({\varphi^{(2)}_{n}\over 2})I-i\sin({\varphi^{(2)}_{n}\over 2})
  {\bm \Upsilon}^{(2)}\cdot{\bm \sigma}\right].
  \label{sincosrotationL}
  \end{eqnarray}
Here,  $e^{i\chi^{(2)}}$ is a pure phase, $\varphi^{(2)}_n$ denotes the rotation angle, and ${\bm \Upsilon}^{(2)}$ denotes  the unit rotation axis generated by a reflection at the second mirror. Substituting Eqs.~$(\ref{QIL})$ and $(\ref{QRL})$ into Eq.~(\ref{xiRL}), the relation between the incident and reflected spin orientations at the second mirror reads
\begin{eqnarray}
\xi_L= {B e^{2ik'_{3n}}\over C} {\cos\Theta\over(ik'_{3n}-m'\cos\Theta)}
\left( E'_n I-ik'_{3n} \tan\Theta\sigma_3\right)
\xi_R .
\label{spinincidentreflectedL}
\end{eqnarray}
By decomposing the two-component spinor, as shown in Eq.~(\ref{xiIR}), the relation in Eq.~(\ref{spinincidentreflectedL}) can be written in an explicit form as
\begin{eqnarray}
 \begin{pmatrix}
\alpha_L\\
\beta_L
\end{pmatrix}
={B e^{2ik'_{3n}}\over C} {\cos\Theta\over(ik'_{3n}-m'\cos\Theta)}
 \begin{pmatrix}
(E'_n-ik'_{3n} \tan\Theta)\alpha_R\\
(E'_n+ik'_{3n} \tan\Theta)\beta_R 
\label{rotationspinL}
\end{pmatrix}.
\end{eqnarray}
Using the equivalence expressions in Eq.~(\ref{xiRL}) with Eqs.~(\ref{sincosrotationL}) and (\ref{spinincidentreflectedL}), we obtain
\begin{eqnarray}
&&e^{i\chi^{(2)}}\cos({\varphi^{(2)}_{n}\over 2})={B e^{2ik'_{3n}}\over C} {E'_n \cos\Theta\over(ik'_{3n}-m'\cos\Theta)},
\label{cosrotL}\\
&&e^{i\chi^{(2)}}\sin({\varphi^{(2)}_{n}\over 2})
{\bm \Upsilon}^{(2)}\cdot{\bm \sigma}={B e^{2ik'_{3n}}\over C} {k'_{3n}\sin\Theta\over(ik'_{3n}-m'\cos\Theta)} \sigma_3.
\label{sinrotL}
\end{eqnarray}
Finally, we obtain the rotation angle and rotation axis at the second mirror as\footnote{ Here, we choose ${\bm \Upsilon}^{(2)}=-\hat z$ because we have chosen ${\bm \Upsilon}^{(1)}=\hat z$ (\ref{rotationangle0}). The opposite direction of the rotation axis is derived from the opposite normal unit vector $N_\mu$ to the first and second mirrors. In addition, if we choose ${\bm \Upsilon}^{(1)}=-\hat z$, we may choose ${\bm \Upsilon}^{(2)}=\hat z$ with the opposite sign of ${\varphi}^{(2)}_n$. } 
  \begin{eqnarray}
  \tan({\varphi^{(2)}_{n}\over 2})=-{k'_{3n}\tan\Theta\over E'_n}~~{\rm and },~~~
  {\bm \Upsilon}^{(2)}=-\hat z
  \label{rotationangleL}.
  \end{eqnarray}

\subsection{Changes of spin orientation at first and second mirrors}
\label{Reflect1st2ndMirrors}

We consider the relation of the spin orientation owing to the reflection at the first and second mirrors. By substituting the components $\alpha_R$ and $\beta_R$ in Eq. (\ref{rotationspin}) into Eq. (\ref{rotationspinL}), we obtain the following relation for consistency:
\begin{eqnarray}
\begin{pmatrix}
\alpha_L\\
\beta_L
\end{pmatrix}
=
e^{i\Xi}
\begin{pmatrix}
\alpha_L\\
\beta_L
\end{pmatrix}.
\label{1and2}
\end{eqnarray}
The phase factor $e^{i\Xi}$ in Eq.~(\ref{1and2}) is explicitly given by
\begin{eqnarray}
e^{i\Xi}={e^{2ik'_{3n}}(-ik'_{3n}-m'\cos\Theta)\over(ik'_{3n}-m'\cos\Theta)},
\label{ConsistentPhase}
\end{eqnarray}
which equals 1 for the allowed momentum (\ref{discretek32}). 
This fact shows that, for a Dirac particle in a 1D box, the reflected spin orientation at the second mirror is completely the same as the incident spin orientation at the first mirror.

 From Eqs.~(\ref{rotationspin}) and (\ref{rotationspinL}), we obtain that the normalized two-component spinor of the left-moving wave component implies $|\alpha_R|^2+|\beta_R|^2=|C|^2/|B|^2$. In other words, the normalization two-component spinor of the right-moving wave component leads to the condition of $|C|^2=|B|^2$ (cf. Refs.~\cite{Menon2004, Alberto11}). Comparing Eqs.~(\ref{rotationangle0}) and (\ref{rotationangleL}), we find that the rotation angle generated by a reflection at the first mirror can be related to the rotation angle generated by the reflection at the second mirror as
\begin{eqnarray}
\varphi^{(2)}_{n}=\varphi^{(1)}_{n},
~~{\rm and }~~ {\bm \Upsilon}^{(2)}=-{\bm \Upsilon}^{(1)}=-\hat z.
\label{rotgeneral}
\end{eqnarray}
From this perspective, although the generated rotational angle at the second mirror is the same as the generated rotation angle at the first mirror, their rotation axes are in opposite directions.  

In general, the rotation angle depends on the momentum, mass, and chiral angle. Next, we consider the behavior of the rotational angles at certain particular limits of $m'$ and specific chiral angles $\Theta$. In the limit $m'\ll 1$, we have the UR limit for Eqs.~(\ref{rotationangle0}) and (\ref{rotationangleL}) as follows:
\begin{eqnarray}
\tan({\varphi^{(1), {\rm UR}}_{n}\over 2})=\tan({\varphi^{(2), {\rm UR}}_{n}\over 2})\simeq -\tan\Theta,
\label{rotangleUR}
\end{eqnarray}
after omitting the higher-order ${\cal O}(m'^2)$. This means that, in the limit $m'\ll 1$, the rotation angles reduce $\varphi^{(1),{\rm UR}}_{n}=\varphi^{(2),{\rm UR}}_{n}=-
2\Theta$.
For the chiral angle $\Theta=0,\pi$, we have  $\varphi^{(1)}_{n}=\varphi^{(2)}_{n}=0$, which means that in 
these cases the spin orientation does not change. 
In the NR limit $m'\gg 1$, we obtain an approximation expression for the rotation angle as
\begin{eqnarray}
\tan({\varphi^{(1), {\rm NR}}_{n}\over 2})=\tan({\varphi^{(2), {\rm NR}}_{n}\over 2})\simeq -{n\pi\over m'} \tan\Theta,
\label{rotangleNR}
\end{eqnarray}
after omitting the higher-order ${\cal O}(1/m'^2)$.

 The results in Eqs. (\ref{rotgeneral})-(\ref{rotangleNR}) give the relations between the reflections at the first mirror $z'=0$ and the second mirror $z'=1$. At the second mirror, we can choose the same rotational angle but with the opposite direction for rotation axis to that at the first mirror. Alternatively, we may choose the opposite sign of rotation angle but with the same direction of rotation axis. These results show that, in the confinement system of a 1D box, the spin orientation of a Dirac particle is consistently reflected for the allowed discrete momentum, as has been shown with Eq.~(\ref{ConsistentPhase}).  In the UR limit (\ref{rotangleUR}), the reflections change the spin orientation as a function of chiral angle in general. In contrast, in the NR limit (\ref{rotangleNR}), the reflections do not change the spin orientation for all chiral angles if we omit the correction of the order of ${\cal O}(1/m')$. In contrast to the UR limit for the discrete momentum, the energy level, and the transition frequency, the UR limit of rotation angle completely depends on the chiral angle. The rotation angle dependence on the chiral angle is also generated by the role of the chiral angle in the boundary conditions depending on the NR and UR limits. 

\section{Probability, normal probability current, and scalar densities}
\label{probabilitydensity}

 In this section, we follow the procedure in our previous study \cite{Rohim} to analyze the probability, normal probability current, and scalar densities along $z^\prime\equiv z/\ell$. In the parameter $z^\prime$, the Dirac particle is confined to a 1D box with a length normalized by $\ell$.
 
\subsection{Probability density}

The probability density for a Dirac particle in a 1D box is given by
\begin{eqnarray}
\rho_{k^\prime_{3n} s} (z^\prime)&=&\bar\psi_{k^\prime_{3n} s}(z^\prime)\gamma^0\psi_{k^\prime_{3n} s}(z^\prime)\nonumber\\
&=& {2|C|^2\over(m^\prime+E^\prime_n)}\left[2E^\prime_n+ ({\cal D}_ne^{-2ik^\prime_{3n}z^\prime}+{\cal D}^*_ne^{2ik^\prime_{3n}z^\prime})m^\prime\right],
\label{probdensityequation}
\end{eqnarray}
 where the coefficient ${\cal D}_n$ is given by
\begin{eqnarray}
{\cal D}_n={(-E^\prime_n\cos\Theta+ik^\prime_{3n}\sin\Theta(|\alpha_L|^2-|\beta_L|^2))\over (ik^\prime_{3n}+m^\prime\cos\Theta)}.
\label{CalD}
\end{eqnarray}
To obtain an explicit expression for the probability density (\ref{probdensityequation}), we used  $|\alpha_L|^2+|\beta_L|^2=1$, $|\alpha_R|^2+|\beta_R|^2=|C|^2/|B|^2(=1)$, and the relation between $B$ and $C$ in Eqs.~(\ref{CoefficientRelations1}) and (\ref{CoefficientRelations2}). From Eqs.~(\ref{probdensityequation}) and (\ref{CalD}), it can be seen that the probability density depends on the chiral angle and spin orientation. This dependence originates from the interference between the left- and right-moving wave components. Here, the coefficient $|C|^2$ is determined by 
\begin{eqnarray}
\int_0^1\rho_{k^\prime_{3n} s} (z^\prime)dz'=1~.
\label{NormalizedProb}
\end{eqnarray}

\begin{figure}[!t]
\centering\includegraphics[width=6.0in]{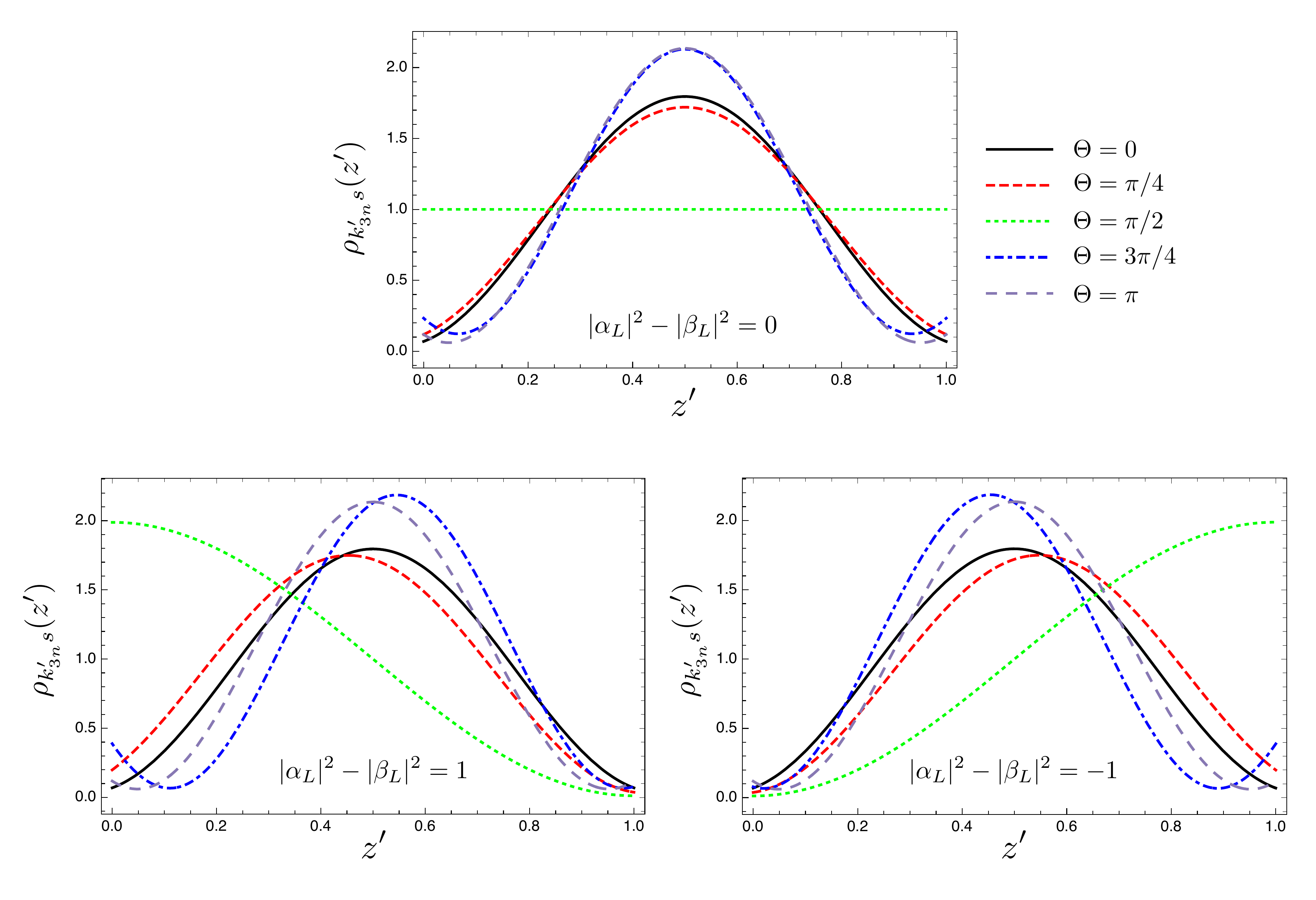}
\caption{ Probability density $\rho_{k^\prime_{3n} s}(z^\prime)$ for the ground state Dirac particle in a 1D box with $m'=10$ and five chiral angles $\Theta=0,\pi/4,\pi/2,3\pi/4,\pi$. We provide three conditions for the spin orientation of the left-moving wave component. The upper panel corresponds to the condition in which the left-moving wave component has a spin orientation in the $\pm x$- or $\pm y$-directions  ($|\alpha_L|^2-|\beta_L|^2=0$), the lower-left panel corresponds to the $+z$-direction  ($|\alpha_L|^2-|\beta_L|^2=1$), and the lower-right panel corresponds to the $-z$-direction  ($|\alpha_L|^2-|\beta_L|^2=-1$). }
\label{Probabilitydensity}
\end{figure}

Figure~\ref{Probabilitydensity} demonstrates the probability density (\ref{probdensityequation}) of a Dirac particle in a 1D box for some values of chiral angles  and some specific spin orientations. As shown above, the probability density generally depends on the chiral angle and spin orientations.  This dependence appears in the second term of the numerator of Eq.~(\ref{CalD}). By contrast, the probability density does not depend on the spin orientations when the  chiral angle $\Theta=0,\pi$ (see  Ref.~\cite{Alberto} for the non-chiral case). This property can be understood from the dynamics of the spin orientation after reflection on the mirror, where the spin orientation does not change when the chiral angle  $\Theta=0,\pi$ for any value of the two-component spinor $\xi_L$, namely, $\alpha_L$ and $\beta_L$. By contrast, for chiral angle $\Theta\neq 0,\pi$, the probability density generally depends on the values of the chiral angle and spin orientation. For example, when the spin orientation of a left-moving wave component is in the $\pm x$- or $\pm y$-directions, i.e., $|\alpha_L|^2-|\beta_L|^2=0$, we have symmetrical shapes that are similar to the behavior of the probability density for the  chiral angle $\Theta=0,\pi$.  By contrast, when the spin orientation of the left-moving wave component is in the $+z$- or $-z$-directions, different properties appear. In these spin orientations, the shape of the probability density becomes asymmetric. The shape of the probability density for a particle with a spin orientation in the $-z$-direction looks like the ``flipping" of the case of the spin orientation in the $+z$-direction. This flipping corresponds to the behavior of the spin orientations around the first and second mirrors. When the left-moving wave component has a spin orientation in the $+z$- or $-z$-direction, its spin orientation does not change because of a reflection for any chiral angle. 
In other words, the spin orientation of the right-moving wave component is the same as the spin orientation of the left-moving wave component, where the value of $|\alpha_L|^2-|\beta_L|^2$ for $+z$-direction has an opposite sign to that of $-z$-direction. This condition may be a source of the asymmetrical distribution of the probability density and further for the scalar density, as we will see below. Because the asymmetrical source is also related to the chiral angle, the asymmetrical behavior does not appear in the probability density for either the NR or the UR limits, where the influence of the chiral angle is quite small.  From Fig.~\ref{Probabilitydensity}, we can also see that when the chiral angle $\Theta=\pi/2$ and the spin orientation is in the $\pm x$- or $\pm y$-directions, we obtain ${\cal D}_n=0$, which means that the probability density is constant along $z'$. This form also appears for the chiral angle $\Theta=3\pi/2$ with the spin orientation in the $\pm x$- or $\pm y$-directions.

\begin{figure}[!t]
\centering\includegraphics[width=4.5in]{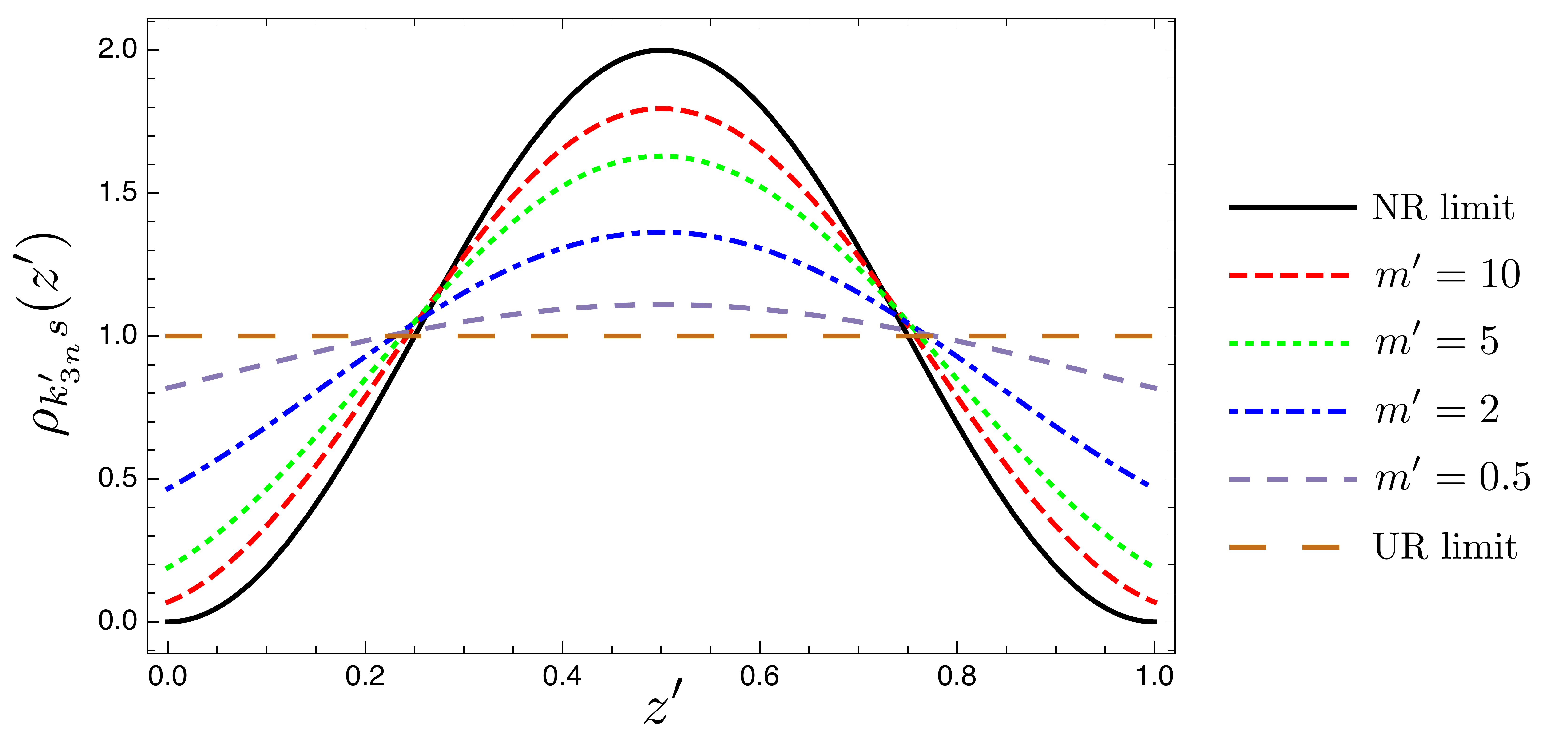}
\caption{Probability density $\rho_{k^\prime_{3n} s}(z^\prime)$ for the ground state Dirac particle in a one-dimensional box with $m' = 0.5, 2, 5, 10$ and the non-chiral case $(\Theta= 0)$ in comparison with NR (\ref{probNR}) and UR (\ref{probUR}).}
\label{Probabilitydensity1}
\end{figure}

 We next consider the probability density for a Dirac particle in a 1D box in the case of the NR limit, for which we have
\begin{eqnarray}
\rho^{\rm NR}_{k'^{\rm NR}_{3n}s}(z')\simeq
2|C|^2\left(1+{\cal F}e^{-2ik'^{\rm NR}_{3n}z'}+{\cal F}^* e^{{2ik'^{\rm NR}_{3n}}z'}\right),
\label{probNR}
\end{eqnarray}
where we have defined 
\begin{eqnarray}
 {\cal F}\simeq{-m'\cos\Theta-ik'^{\rm NR}_{3n}+ik'^{\rm NR}_{3n}\sin\Theta(|\alpha_L|^2-|\beta_L|^2)\over 2m'\cos\Theta},
 \label{calF}
\end{eqnarray}
and omitted the contribution of ${\cal O}(1/m'^{2})$ and its higher order. If we omit the contribution of ${\cal O}(1/m')$ in Eqs.~(\ref{probNR}) and (\ref{calF}), the probability density (\ref{probNR}) reduces to the well-known probability density of NR particles, i.e., in proportion to $\sin^2(k'^{\rm NR}_{3n}z')$ with $k'^{\rm NR}_{3n}=n\pi$. 

In the UR limit $m'\ll1$, we obtain an approximate expression for the probability density as follows:
\begin{eqnarray}
\rho^{\rm UR}_{k'^{\rm UR}_{3n}s}(z')\simeq
{2|C|^2\over k'^{\rm UR}_{3n}}
\left(2 k'^{\rm UR}_{3n}- 2m'+{\cal G}e^{-2ik'^{\rm UR}_{3n}z'}+{\cal G}^* e^{{2ik'^{\rm UR}_{3n}}z'}\right),
\label{probUR}
\end{eqnarray}
where 
\begin{eqnarray}
{ \cal G}\simeq im'\cos\Theta+m'\sin\Theta(|\alpha_L|^2-|\beta_L|^2).
 \label{calG}
\end{eqnarray}
In addition, we omitted the contribution of ${\cal O}(m'^{2})$ and its higher order. If we omit the contribution of ${\cal O}(m')$ and its higher order, the probability density (\ref{probUR}) under a state of $n$ will give a constant value along $z'$, where the discrete momentum becomes  $k'^{\rm UR}_{3n}=(2n-1)\pi/2$. 

 Figure~\ref{Probabilitydensity1} demonstrates the probability density of the ground state Dirac particle in a 1D box for the non-chiral case ($\Theta=0$) and four values for the parameter $m'=0.5,2,5,10$ in comparison with the NR and UR limits. From this figure, it can be seen that the curves become close to that of the NR limit as $m'$ becomes larger than unity. It is also clear that the NR limit leads to a vanishing probability density around the mirrors, which is a natural property of NR particles associated with Dirichlet boundary conditions. By contrast, the curves become close to the UR limit as the parameter $m'$ becomes smaller, where the probability density approaches a constant distribution, which appears as a straight horizontal line. This is because the probability density is dominated by its first term on the right-hand side of Eq.~(\ref{probUR}) in the UR limit. The correction is of the order of $m/k_{3n}$.

The authors of Ref.~\cite{Alberto} demonstrated the probability density of a Dirac particle parameterized by the box's size as a function of Compton wavelength. Figure~\ref{Probabilitydensity1} shows the same properties as those of Fig. 5 of Ref.~\cite{Alberto}, which is obtained for the ground state. The larger box puts  the system close to the NR limit. 
In contrast, the smaller box's size make the system close to the UR limit.

\subsection{Normal probability current density}

The normal probability current density is expressed as
\begin{eqnarray}
J_{N, k'_{3n}s}(z^\prime)=
J^{3}_{k^\prime_{3n} s}(z^\prime)=\bar\psi_{k^\prime_{3n} s}(z^\prime)\gamma^3\psi_{k^\prime_{3n}s}(z^\prime). 
\label{Normalprobability}
\end{eqnarray}
Here, we adopt the same strategy with the probability density, i.e.,   $|\alpha_L|^2+|\beta_L|^2=1$, $|\alpha_R|^2+|\beta_R|^2=|C|^2/|B|^2(=1)$, and use the relation between $B$ and $C$ in Eqs.~(\ref{CoefficientRelations1}) and (\ref{CoefficientRelations2}).  For a Dirac particle in a 1D box, the contribution of the left-moving wave component to the normal probability current density is canceled by the contribution of the right-moving wave component. The interference term between the left- and right-moving wave components vanishes everywhere. Finally, we find that the total normal probability density vanishes everywhere as well  (cf. Refs.~\cite{Boulanger2006, Rohim}). This condition indicates that we cannot use the vanishing normal probability current density at the boundary surface as the boundary condition for our system.

\begin{figure}[!t]
\centering\includegraphics[width=6.0in]{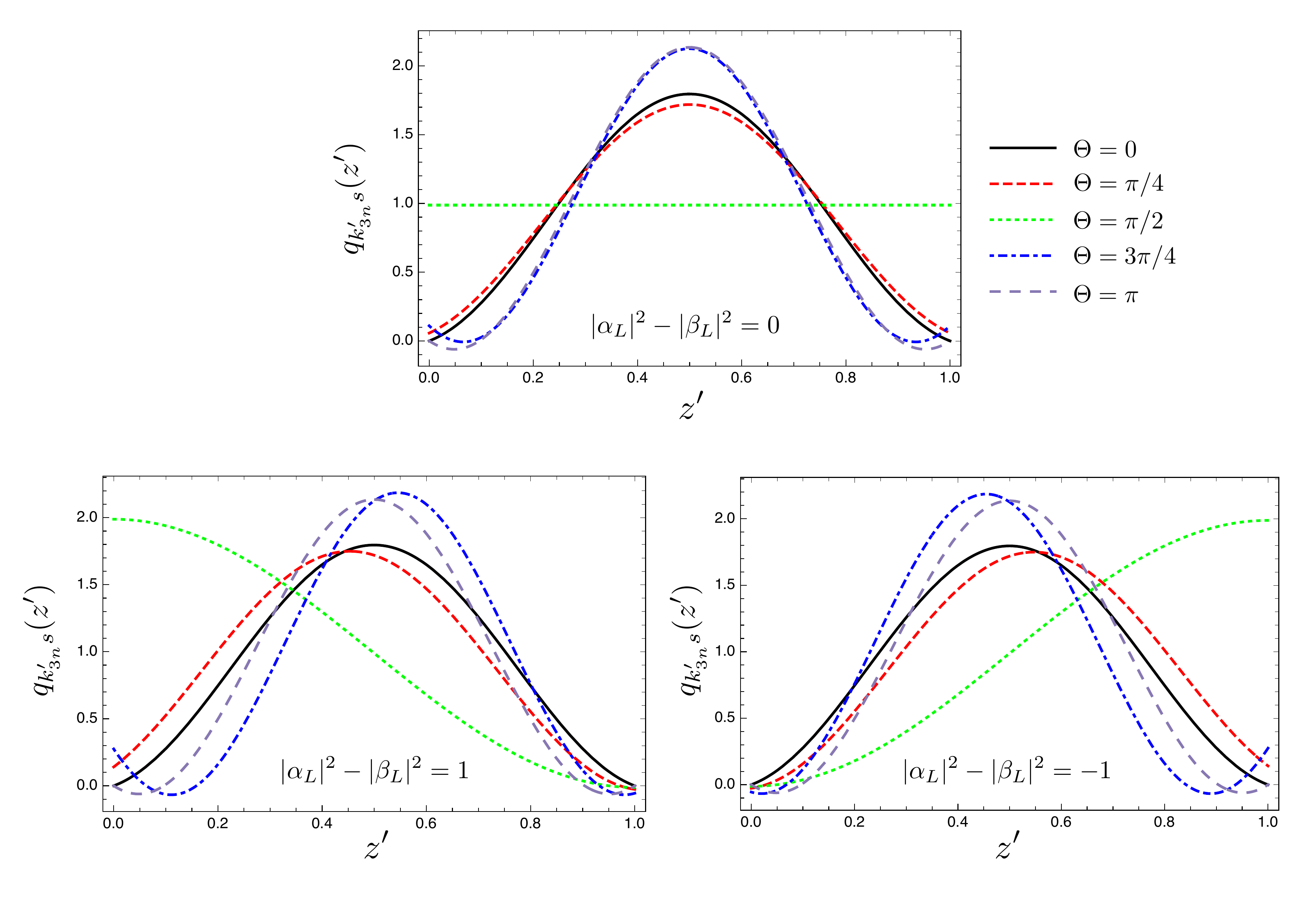}
\caption{Scalar density $q_{k^\prime_{3n} s}(z^\prime)$ for the ground state Dirac particle in a 1D box.
Here, we use the same parameters for the chiral angle and spin orientation as in Fig.~\ref{Probabilitydensity} for each panel.}
\label{FigScalarDensity}
\end{figure} 

\subsection{Scalar density}

The scalar density for a Dirac particle in a 1D box is given by
\begin{eqnarray}
q_{k^\prime_{3n} s} (z^\prime)&=&\bar\psi_{k^\prime_{3n} s}(z^\prime)\psi_{k^\prime_{3n} s}(z^\prime)\nonumber\\
&=& {2|C|^2\over(E^\prime_n+m^\prime)}\left[2m^\prime+ ({\cal D}_ne^{-2ik^\prime_{3n}z^\prime}+{\cal D}^*_ne^{2ik^\prime_{3n}z^\prime})E^\prime_n\right],
\label{ScalarDensityEq}
\end{eqnarray}
where ${\cal D}_n$ is given by Eq.~(\ref{CalD}).  Similar to the probability and normal probability current densities, to obtain the scalar density, herein we  have used $|\alpha_L|^2+|\beta_L|^2=1$, $|\alpha_R|^2+|\beta_R|^2=|C|^2/|B|^2(=1)$, and the relation between $B$ and $C$ in Eqs.~(\ref{CoefficientRelations1}) and (\ref{CoefficientRelations2}).  Here, the coefficient $|C|^2$ is determined by Eq.~(\ref{NormalizedProb}).

Figure~\ref{FigScalarDensity} plots the curves of scalar density for the ground state Dirac particle in a 1D box for $m'=10$ with various chiral angles and spin orientation. The curves show that the scalar density vanishes at the boundary surface when the chiral angle $\Theta=0,\pi$. This behavior is consistent with the default properties of BC-chiral MIT, which guarantees a vanishing scalar density around the mirror for the chiral angle $\Theta=0, \pi$; see Eq.~(\ref{probandscalardensities}). In addition, at these chiral angles, the scalar density does not depend on the spin orientation. Similar to the probability density, the scalar density for a Dirac particle in a box depends on the chiral angle and spin orientation in general. For example, when the left-moving wave component has a spin orientation in the $\pm x$- or $\pm y$-directions, the distribution of scalar density is symmetrical, whereas when the left-moving wave component has a spin orientation in the $+z$- or $-z$-directions, the distribution of the scalar density is asymmetrical. Similar to the probability density, this different behavior corresponds to the rotation or changes in the spin orientation owing to reflections from the mirrors, which depend on the chiral angle and spin orientation.

\begin{figure}[!t]
\centering\includegraphics[width=4.5in]{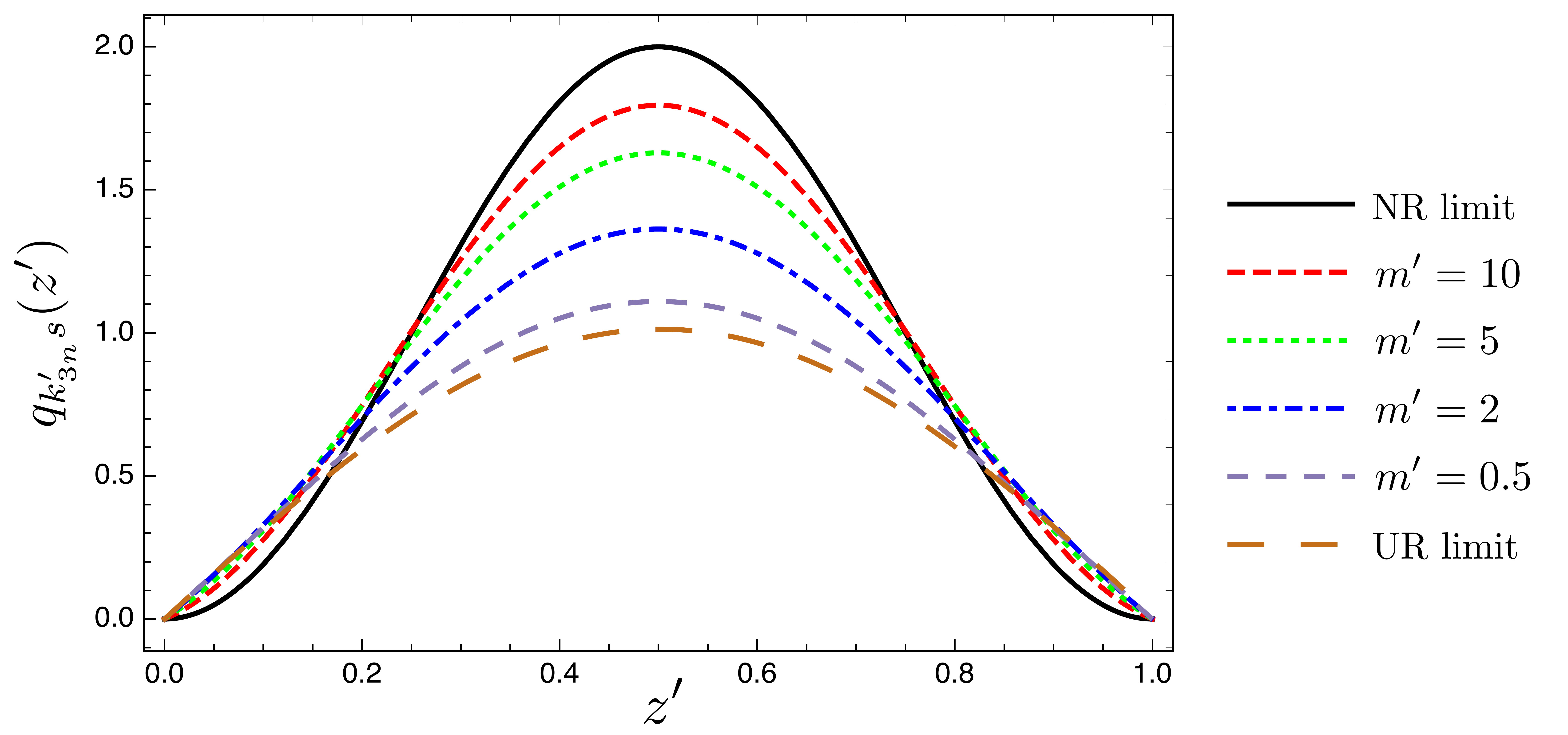}
\caption{Scalar density $q_{k^\prime_{3n} s}(z^\prime)$ for the ground state Dirac particle in a one-dimensional box. Here, we used the same parameters as those in  Fig.~\ref{Probabilitydensity1}. 
} 
\label{Scalardensity1}
\end{figure}

Let us consider the scalar density at a special limit. In the UR limit,  the approximation form for the scalar density is
\begin{eqnarray}
q^{\rm UR}_{k^{\prime{\rm UR}}_{3n} s} (z^\prime)\simeq {2|C|^2\over k^{\prime{\rm UR}}_{3n}}
\left( 2m'
+{\cal P}e^{-2ik^{\prime{\rm UR}}_{3n}z'}+{\cal P}^*e^{2ik^{\prime{\rm UR}}_{3n}z'}
\right),
\end{eqnarray}
where
\begin{eqnarray}
 {\cal P}\simeq (im'\cos\Theta+k^{\prime{\rm UR}}_{3n}-m')(i\cos\Theta +\sin\Theta (|\alpha_L|^2-|\beta_L|^2),
\end{eqnarray}
and we have omitted the contribution of the order ${\cal O}(m'^{2})$ and its higher order. In the NR limit, the approximation form for the scalar density is
\begin{eqnarray}
q^{\rm NR}_{k^{\prime{\rm NR}}_{3n} s} (z^\prime)\simeq 2|C|^2
\left(1
+{\cal F}e^{-2ik^{\prime{\rm NR}}_{3n}z'}+{\cal F}^*e^{2ik^{\prime{\rm NR}}_{3n}z'}
\right),
\label{scalarNR}
\end{eqnarray}
where ${\cal F}$ is given by Eq.~(\ref{calF}), which is shared by the probability density of the NR limit. In this limit, we omit the contribution of the order ${\cal O}(1/m'^{2})$ and its higher order. 

Figure~\ref{Scalardensity1} provides the curves of the scalar density for the ground state Dirac particle in a 1D box. We used the non-chiral case ($\Theta=0$) with four values of $m'=0.5,2,5,10$, which we used in Fig.~\ref{Probabilitydensity1} to compare with the  NR and UR limits. The curves show similar behavior of vanishing scalar density at the boundary. The scalar density vanishing at the boundary surface is a consequence of the non-chiral case or BC-MIT.

Comparing the curves of probability densities in Fig. \ref{Probabilitydensity} with the scalar densities in Fig.~\ref{FigScalarDensity}, they look quite similar. This is because we used the parameter of $m'=10$, which is relatively close to the NR limit \cite{Alberto, Alberto11} (see also Figs.~\ref{Probabilitydensity1} and \ref{Scalardensity1}). However, for some values of the chiral angle, their behavior around the boundary surface is different. The probability density for the relativistic particle under BC-chiral MIT does not vanish at the boundary surface. In contrast, its scalar density vanishes when the chiral angle takes the values of $\Theta=0,\pi$, as has been shown in Eq.~(\ref{probandscalardensities}). For the case $m'=10$, the order of the difference between probability and scalar densities is relatively small. If we choose the smaller values of parameter $m'$, the difference becomes significant. In the NR limit, both the probability and scalar densities show a similar curve (see Eqs.~(\ref{probNR}) and  (\ref{scalarNR})). However, the curve of probability density in the UR limit differs from its scalar density in the same limit. The behavior of the UR limit of probability density for the non-chiral case is similar to the relativistic probability density of the case $\Theta={\pi\over 2}$, which becomes constant along the $z'$-axis, whereas for the scalar density in the UR limit of the case $\Theta=0$, such a property does not appear. Its behavior inside the box for $m'=0$ and $\Theta=0$ is described by the sine function, $\sin (2k^{'{\rm UR}}_{3n} z')$, rather than a constant along the $z'$-axis.

\section{Summary and conclusion}
\label{Summary}

We studied the effects of the boundary conditions for a Dirac particle in a 1D box. We adopt BC-chiral MIT \cite{Lutken}, which is a generalized form of BC-MIT through the introduction of a chiral angle. This study is a generalization of the work of the authors of Ref.~\cite{Alberto}, which investigated Dirac particle in a box using BC-MIT. 
Using BC-chiral MIT, we found generalized bound states, which reduce to the previous results in the non-chiral case in Ref.~\cite{Alberto}. We have also shown that the discrete momentum, energy level, and transition frequency depend on the chiral angle but do not depend on the spin orientation. Compared with the non-chiral case, the values of the discrete momentum,  energy level, and transition frequency can be lower or higher depending on the values of the chiral angle. Following the procedure in Ref.~\cite{Nicolaevici}, we investigated how the spin orientation changes around the boundaries under BC-chiral MIT. The spin state is in a consistent state to repeat the reflections at both boundaries. 

We also demonstrated the role of the chiral angle in the probability, normal probability current, and scalar densities under BC-chiral MIT. In general, the probability and scalar densities depend on the chiral angle and spin orientation. For the case $\Theta=0,\pi$, the probability and scalar densities do not depend on the spin orientation. However, for the cases of $\Theta\neq0,\pi$, they do. At the boundary, the probability density does not vanish. Similar to the probability density, the scalar density depends on the chiral angle. For the cases in which $\Theta=0,\pi$, it vanishes at the boundaries. We also found that the probability and scalar densities can be asymmetric depending on the spin orientation and chiral angle. The normal probability current density for a Dirac particle in a 1D box vanishes everywhere. 

We investigated the NR and UR limits for the discrete momentum, energy level, transition frequency, probability density, and scalar density in an analytic manner. Under these limits, the results show that the contribution of the chiral angle is coupled to the parameter $m'$ and that the influence is quite small. In other words, the contribution of the chiral angle is negligible in these limits. However, when the chiral angle takes the values $\Theta=\pi/2,3\pi/2$ for any value parameter $m'$, the discrete momentum is not trivial. For these chiral angles, the discrete momentum does not depend on parameter $m'$.

We clarified the basic properties of wavefunctions of a Dirac field in a box with BC-chiral MIT. This should be useful in studying the finite volume effect associated with Dirac fermions. For example, the Nambu–Jona-Lasinio model in a finite volume region with a generalized boundary condition would be an interesting application; this is left for future research (cf.~Refs.~\cite{Chernodub2016,Chernodub}).

\section*{Acknowledgment}

We thank Y. Kojima and N. Okabe for useful discussions. 
We also thank S.-Y. Lin, K. Ueda, and A. Matsumura for discussions related to this study. A.R. thanks the member of the Theoretical Astrophysics Laboratory of Kyushu University, where most of this study was conducted, for the hospitality provided during his research visit.  A.R. is supported by  a Japanese Government (Monbukagakusho: MEXT) Scholarship.

\appendix
 
 \section{Dirac particle in Minkowski coordinates and its spin orientation}
 \label{DiracMinkownskiSolution}
 
\subsection{Dirac wavefunction in Minkowski coordinates}

The Dirac equation in Minkowski coordinates is given by 
\begin{eqnarray}
(i\gamma^\mu\partial_\mu-m)\Psi_{{\bm k}s}(t,{\bm x})=0,
\label{DiracEquation}
\end{eqnarray}
where $m$ is the mass of the Dirac particle, ${\bm k}$ is the momentum, and $s$ denotes the spin orientation. Here,  $\gamma^\mu$ are gamma matrices, which are explicitly written in a Dirac representation as
\begin{eqnarray}
\gamma^0=
\left(\begin{array}{cccc}
I&0 \\
0&-I
\end{array}\right)~,~~ 
\gamma^i=
\left(\begin{array}{cccc}
0&\sigma_i \\
-\sigma_i&0
\end{array}\right)~ ,~~
i=1,2,3.
\label{DiracRepresentation}
\end{eqnarray} 
Note that here  $I$ is a $2\times2$ identity matrix, and $\sigma_i$ are Pauli matrices, which satisfy anti-commutation relations $\{\sigma_i,\sigma_j\}=2\delta_{ij}I$. The gamma matrices satisfy $\lbrace\gamma^\mu,\gamma^\nu\rbrace=2\eta^{\mu\nu}$, where we use the notation  
$\eta^{\mu\nu}={\rm diag}(1,-1,-1,-1)$ in the present paper. 
To find the Dirac wavefunction, we introduce the ansatz for a positive frequency solution as 
\begin{eqnarray}
\Psi_{{\bm k}s}(t,{\bm x})=u_{{\bm k}s} e^{-iE t}e^{i{\bm k}\cdot{\bm x}},
\end{eqnarray}
 where  $E= \sqrt{m^2+|{\bm k}|^2}$ is the energy of the Dirac particle and $u_{{\bm k}s}$ is a  four-component Dirac spinor. Then, the spinor  $u_{{\bm k}s}$ can be decomposed into the upper two-component spinor $\xi_s$ and lower two-component spinor $\chi_s$ as 
 \begin{eqnarray}
 u_{{\bm k}s}=  
 \begin{pmatrix}
\xi_s\\
\chi_s
\end{pmatrix}.
\label{Diracspinor}
 \end{eqnarray} 
Substituting back the Dirac spinor (\ref{Diracspinor}) into the Dirac equation (\ref{DiracEquation}), we obtain the following two coupled equations
 \begin{eqnarray}
(E-m)\xi_s- {\bm \sigma}\cdot{\bm k}\chi_s=0,
 \label{NegativeEnergy}\\
(m+E)\chi_s-{\bm \sigma}\cdot{\bm k}\xi_s=0,
\label{PositiveEnergy}
 \end{eqnarray}
 where Eqs.~(\ref{NegativeEnergy}) and  (\ref{PositiveEnergy})  are associated with negative energy and positive energy solutions, respectively. Herein, we focus on the solution for positive energy, which implies that the Dirac solution  can be written explicitly as
 \begin{eqnarray}
\Psi_{{\bm k} s}(t,{\bm x})=
 {\cal N}_{{\bm k}s}
  \begin{pmatrix}
\xi_s\\
\frac{\bm \sigma\cdot{\bm k}}{(m+E)}\xi_s
\end{pmatrix} e^{-iE t}e^{i{\bm k}\cdot{\bm x}},
\label{Diracsolution}
 \end{eqnarray}
 where the two-component spinor $\xi_s$ is normalized by the condition of  $\xi_s^\dagger\xi_s=1$ and ${\cal N}_{{\bm k}s}$ is the normalization constant obtained by
  \begin{eqnarray}
(\Psi_{{\bm k}s}, \Psi_{{\bm k}^\prime s^\prime})=
 \int d^3{\bm x} \Psi_{{\bm k}s}^{\dagger}\Psi_{{{\bm k}'s'}}
 = \delta({\bm k}-{\bm k^\prime})\delta_{s s^\prime}.
 \label{NormalizationConstantss}
 \end{eqnarray}

 \subsection{Spin orientation}
\label{spincomponent}

In general, the two-component spinor $\xi_s$ consists of a linear combination of a two-component spinor with a spin-up 
($+z$-direction) $\xi_{+z}=(1,0)^{\rm T}$  and a spin-down ($-z$-direction)  
$\xi_{-z}=(0,1)^{\rm T}$ as 
\begin{eqnarray}
\xi_s=\alpha\xi_{+z} + \beta \xi_{-z}
=
\begin{pmatrix}
\alpha\\
\beta
\end{pmatrix}
,
\end{eqnarray}
where $|\alpha|^2+|\beta|^2=1$.  
The normalized two-component spinors with spin orientation in the $\pm x$- and $\pm y$-directions are given by
\begin{eqnarray}
&&\xi_{+x}=
  {1\over \sqrt{2}}
  \begin{pmatrix}
1 \\
1
\end{pmatrix}, ~~
\xi_{-x}=
{1\over \sqrt{2}}
  \begin{pmatrix}
1 \\
-1
\end{pmatrix},\\
&&\xi_{+y}=
{1\over \sqrt{2}}
  \begin{pmatrix}
1 \\
i
\end{pmatrix}, ~~
\xi_{-y}=
{1\over \sqrt{2}}
  \begin{pmatrix}
1 \\
-i
\end{pmatrix}.
\end{eqnarray}
The spin orientation can also be parameterized through the polar angle $\theta$ and azimuthal angle $\phi$ as follows:
\begin{eqnarray}
\begin{pmatrix}
\alpha \\
\beta
\end{pmatrix}
=
\begin{pmatrix}
e^{-i\phi/2} \cos(\theta/2)\\
e^{i\phi/2} \sin(\theta/2)
\end{pmatrix}
.
\end{eqnarray}

\let\doi\relax


\begin{thebibliography}{9}

 \bibitem{Alberto2017}
P. Alberto, S. Das, and E. C. Vagenas, Eur. J. Phys. \textbf{39}, 025401 (2018).

\bibitem{Alkhateeb}
M.~Alkhateeb and A.~Matzkin, arXiv:2103.06538.



 
\bibitem{Alberto}
P. Alberto, C. Fiolhais, and V. M. S. Gil, Eur. J. Phys. {\bf 17}, 19 (1996).
  
\bibitem{Alonso1997}
V. Alonso and S. De Vincenzo, J. Phys. A: Math. Gen. {\bf 30}, 8573 (1997). 
  


\bibitem{Menon2004}
G.~Menon and S. Belyi, Phys. Lett. A {\bf 330}, 33 (2004).
  
\bibitem{Toyama2010a}
F. M. Toyama and Y. Nogami, Phys. Rev. A {\bf 81}, 044106 (2010).
  
\bibitem{Toyama2010}
F.~M.~Toyama and Y.~Nogami, Phys. Lett. A \textbf{374}, 3838 (2010).

  
\bibitem{Alberto11}
P. Alberto, S. Das, and E. C. Vagenas, Phys. Lett. A \textbf{375}, 1436 (2011). 

\bibitem{AlHashimi2011}
M. H. Al-Hashimi and U. J. Wiese, Ann. Phys. \textbf{327}, 1 (2012).

\bibitem{Sitenko2014}
Y. A. Sitenko, Phys. Rev. D \textbf{91}, 085012 (2015). 

\bibitem{AlHashimi}
M. H. Al-Hashimi, A. M. Shalaby, and U. J. Wiese, Phys. Rev. D \textbf{95}, 065007 (2017).


\bibitem{Chodos1}
A.~Chodos, R.~L.~Jaffe, K.~Johnson, C.~B.~Thorn, and V.~F.~Weisskopf, Phys. Rev. D \textbf{9}, 3471 (1974).

\bibitem{Chodos2}
A.~Chodos, R.~L.~Jaffe, K.~Johnson, and C.~B.~Thorn, Phys. Rev. D \textbf{10}, 2599 (1974).

\bibitem{Johnson}
K.~Johnson, Acta Phys. Polon. B \textbf{6}, 865 (1975).

  
\bibitem{AWThomas}
A. W. Thomas, Adv. Nucl. Phys. A {\bf 13}, 1 (1984).
 
\bibitem{Chodos1975}
A. Chodos and C. B. Thorn, Phys. Rev. D {\bf 12}, 2733 (1975).


\bibitem{Brown1979}
G.~E.~Brown and M.~Rho, Phys. Lett. B \textbf{82}, 177  (1979).


\bibitem{Hosaka1996}
A. Hosaka and H. Toki, Phys. Rept. \textbf{277}, 65 (1996).

\bibitem{Holt2014}
J.~W.~Holt, M.~Rho, and W.~Weise, Phys. Rept. \textbf{621}, 2 (2016).


\bibitem{Theberge1980}
S. Th\'{e}berge, A. W. Thomas, and G. A. Miller, Phys. Rev. D {\bf 22}, 2838 (1980), Phys. Rev. D {\bf 23}, 2106 (1981)[erratum]. 


 \bibitem{Lutken}
C. A. L\"utken and F. Ravndal, J. Phys. G: Nucl. Phys. \textbf{10}, 123 (1984).


\bibitem{Jaffe}
R. L. Jaffe and A. Manohar, Ann. Phys. \textbf{192}, 321 (1989).




\bibitem{AlHashimi2012}
M. H. Al-Hashimi and U. J. Wiese, Ann. Phys. \textbf{327}, 2742 (2012).

\bibitem{AlHashimi2015}
M. H. Al-Hashimi, A. M. Shalaby, and U. J. Wiese, Ann. Phys. \textbf{362}, 621 (2015).


\bibitem{Chernodub}
M.~N.~Chernodub and S.~Gongyo, Phys. Rev. D \textbf{95}, 096006 (2017).

\bibitem{Ambrus2015}
V.~E.~Ambrus and E.~Winstanley, Phys. Rev. D \textbf{93}, 104014 (2016).

\bibitem{Chernodub2016}
M.~N.~Chernodub and S.~Gongyo, J. High. Energy Phys. \textbf{1701}, 136 (2017).

\bibitem{Chernodub2017}
M.~N.~Chernodub and S.~Gongyo, Phys. Rev. D \textbf{96}, 096014 (2017).

\bibitem{Nicolaevici}
N.~Nicolaevici, Eur. Phys. J. Plus \textbf{132}, 21 (2017). 

\bibitem{Zhang2019}
Z. Zhang, C. Shi, and H. Zong, Phys. Rev. D \textbf{101}, 043006 (2020).

\bibitem{Jenke2011}
T.~Jenke, P.~Geltenbort, H.~Lemmel, and H.~Abele, Nat. Phys. \textbf{7}, 468  (2011).

\bibitem{Rohim}
A.~Rohim, K.~Ueda, K.~Yamamoto, and S.-Y.~Lin, Int. J. Mod. Phys. D {\bf 30}, 2150098 (2021).

\bibitem{Boulanger2006}
N.~Boulanger, P.~Spindel, and F.~Buisseret, Phys. Rev. D \textbf{74}, 125014 (2006).

\end{thebibliography}
\end{document}